\documentclass[11pt]{amsart}
\usepackage[dvips,colorlinks=true,linkcolor=blue]{hyperref}
\numberwithin{equation}{section}
\def\one{\leavevmode\hbox{\small1\kern-3.35pt\normalsize1}}
\newcommand{\car}{\mathbf{1}}
\newcommand{\R}{\mathbb{R}}

\newcommand{\T}{\mathbb{T}}

\newcommand{\Z}{\mathbb{Z}}

\newcommand{\N}{\mathbb{N}}

\newcommand{\C}{\mathbb{C}}

\newcommand{\equ}{\operatornamewithlimits{\sim}}
\newcommand{\D}{\displaystyle}

\newcommand{\Sc}{{\mathcal S}}
\newcommand{\Coi}{{\mathcal C}_0^{\infty}}
\newcommand{\esp}{\mathbb{E}}

\newcommand{\Tr}{\text{tr}}

\newcommand{\F}{{\mathcal F}}
\theoremstyle{plain}
\newtheorem{Th}{Theorem}[section]
\newtheorem{Le}{Lemma}[section]
\newtheorem{Pro}{Proposition}[section]

\theoremstyle{definition}
\newtheorem{Rem}{Remark}[section]
\newtheorem{Def}{Definition}[section]
\title{The band-edge behavior of the density of surfacic states}
\author{Werner Kirsch}
\address[Werner Kirsch]{Fakult{\"a}t f{\"u}r Mathematik and SFB-TR 12, Ruhr Universit{\"a}t
  Bochum, D-44780 Bochum, Deutschland}
\email{\href{mailto:werner@mathphys.ruhr-uni-bochum.de}{werner@mathphys.ruhr-uni-bochum.de}}
\author{Fr{\'e}d{\'e}ric Klopp}
\address[Fr{\'e}d{\'e}ric Klopp]{LAGA, U.M.R. 7539 C.N.R.S, Institut Galil{\'e}e,
  Universit{\'e} de Paris-Nord, 99 venue J.-B. Cl{\'e}ment, F-93430
  Villetaneuse, France}
\email{\href{mailto:klopp@math.univ-paris13.fr}{klopp@math.univ-paris13.fr}}
\keywords{}
\subjclass{}
\begin{document}
\begin{abstract}
  This paper is devoted to the asymptotics of the density of surfacic
  states near the spectral edges for a discrete surfacic Anderson
  model.  Two types of spectral edges have to be considered :
  fluctuating edges and stable edges. Each type has its own type of
  asymptotics.  In the case of fluctuating edges, one obtains Lifshitz
  tails the parameters of which are given by the initial operator
  suitably ``reduced'' to the surface. For stable edges, the surface
  density of states behaves like the surface density of states of a
  constant (equal to the expectation of the random potential) surface
  potential. Among the tools used to establish this are the
  asymptotics of the surface density of states for constant surface
  potentials.
                                %
%%   \vskip.5cm
%%   \par\noindent  \textsc{R{\'e}sum{\'e}.}
%%                                 %
%%   Dans ce travail, nous {\'e}tablissons les asymptotiques de la densit{\'e}
%%   d'{\'e}tats surfaciques pour un mod{\`e}le d'Anderson surfacique discret au
%%   bord du spectre. Pour ce faire, nous distinguons deux types de bord
%%   de spectre, les bords fluctuants et ceux stables. Chaque type de
%%   bord admet des asymptotiques qui lui sont propres. Dans le cas des
%%   bords fluctuants, on obtient des asymptotiques de Lifshitz dont les
%%   param{\`e}tres sont calcul{\'e}s {\`a} partir de l'op{\'e}rateur de d{\'e}part r{\'e}duit au
%%   bord. Pour les bords stables, nous obtenons que la densit{\'e} d'{\'e}tats
%%   surfacique se comporte comme la densit{\'e} d'{\'e}tat surfacique pour un
%%   potential constant {\'e}gal {\`a} l'esp{\'e}rance du potentiel al{\'e}atoire le long
%%   de la surface. La preuve de ce r{\'e}sultat passe par le calcul des
%%   asymptoiques de la densit{\'e} d'{\'e}tats surfaciques pour des potentiels
%%   surfaciques constants.
%
\end{abstract}
\setcounter{section}{-1}
\maketitle
\section{Introduction}
\label{sec:introduction}
On $\Z^d$ ($d=d_1+d_2$, $d_1>0$, $d_2>0$), we consider random
Hamiltonians of the form
\begin{equation*}
  H_\omega=-\frac12\Delta+V_\omega
\end{equation*}
where
\begin{itemize}
\item $-\Delta$ is the free Laplace operator, i.e., $-(\Delta
  u)(n)=\sum_{|m-n|=1}u(m)$;
\item $V_\omega$ is a random potential concentrated on the sub-lattice
  $\Z^{d_1}\times \{0\}\subset\Z^d$ of the form
\begin{equation}
\label{eq:V}
  V_\omega(\gamma_1,\gamma_2)=\begin{cases}\omega_{\gamma_1}&\text{ if
   }\gamma_2=0,\\  0&\text{ if }\gamma_2\not=0.\end{cases},
   \gamma=(\gamma_1,\gamma_2)\in\Z^{d_1}\times \Z^{d_2}=\Z^d.
\end{equation}
and $(\omega_{\gamma_1})_{\gamma_1\in\Z^{d_1}}$ is a family of i.i.d.
bounded random variables. For the sake of simplicity, let us assume
that the random variables are uniformly distributed in $[a,b]$
($a<b$).
\end{itemize}
To keep the exposition as simple as possible in the introduction, we
use these quite restrictive assumptions. We will deal with more
general models in the next section.
\par The operator $H_\omega$ is bounded for almost every $\omega$. It is
ergodic with respect to shifts parallel to the surface. So we know there exists $\Sigma$
the almost sure spectrum of $H_\omega$ (see
e.g.~\cite{Ki:89,MR94h:47068}.
\par For $H_\omega$, one defines the integrated density of surface
states (the IDSS in the sequel), in the following way (see
e.g.~\cite{englisch88:_densit,MR2000j:81290,MR2001b:81026,MR1770547}):
for $\varphi\in\Coi(\R)$, we set
\begin{equation}
  \label{eq:11}
  (\varphi,n_s)=\esp(\Tr(\Pi_1[\varphi(H_\omega)-\varphi(-\frac12\Delta)]\Pi_1))
\end{equation}
where $\Pi_1$ is the orthogonal projector on the subspace
$\C\delta_0\otimes\ell^2(\Z^{d_2})\subset\ell^2(\Z^d)$. Here,
$\delta_0$ denotes the vector with components
$(\delta_{0j})_{j\in\Z^{d_1}}$.
\par Obviously, equation~\eqref{eq:11} defines the integrated density
of surface states $n_s$ only up to a constant. We choose this constant
so that $n_s$ vanishes below $\Sigma\cup\Sigma_0$ where $\Sigma_0$ is
the spectrum of $-\frac12\Delta$. We will see later on that, up to addition
of a well controlled distribution, $n_s$ is a positive measure.
\par One knows that $\Sigma=\sigma(-\frac12\Delta)\cup$supp$(dn_s)$
(see~\cite{englisch88:_densit,MR92a:47058,MR2000j:81290}. We will
study the behavior of $n_s$ at the edges of $\Sigma$. To simplify this
set as much as possible, we will assume that the support of the random
variables $(\omega_{\gamma_1})_{\gamma_1\in\Z^{d_1}}$ is connected.
Under this assumption, we know that
\begin{Le}
  \label{le:1}
  $\Sigma$ is a compact interval given by
  \begin{equation}
    \label{eq:83}
    \Sigma=\sigma(-\frac12\Delta_{d_1})+\bigcup_{\omega_0\in[a,b]}
    \sigma(-\frac12\Delta_{d_2}+\omega\Pi^2_0)
  \end{equation}
  where $\Pi_0^2$ is the projector on the unit vector
  $\delta^2_0\in\ell^2(\Z^{d_2})$.
\end{Le}
This is a consequence of a standard characterization of $\Sigma$ in
terms of periodic potentials (see~\cite{Ki:89,MR94h:47068}). The
assumption that the random variables have connected support can be
relaxed; more connected components for the support of the random
variables will in general give rise to more spectral edges (as in the
case of bulk randomness, see~\cite{Kl:97b}). For the value of
$\Sigma$, two different possibilities occur :
\begin{enumerate}
\item $\Sigma=\sigma(-\frac12\Delta)+[-\alpha,\beta]=[-d-\alpha,d+\beta]$
  where $\alpha=\alpha(a)$, $\beta=\beta(b)$ and $\alpha+\beta>0$;
  this occurs
  \begin{itemize}
  \item if $d_2\leq 2$ and either $a<0$, in which case $\alpha(a)>0$,
    or $b>0$, in which case $\beta(b)>0$,
  \item if $d_2\geq 3$ and $a>a_0$ or $b>b_0$, where,
    by~\eqref{eq:83}, the thresholds $a_0$ and $b_0$ are uniquely
    determined by the family of operators
    $(-\frac12\Delta_{d_2}+t\Pi^2_0)_{t\in\R}$.
  \end{itemize}
  If $\alpha>0$ (resp. $\beta>0$), we say that the left (resp. right)
  edge is a ``fluctuation edge'' or ``fluctuation boundary''
  (see~\cite{MR94h:47068}). If $\alpha=0$ (resp. $\beta=0$), we will
  speak of a ``stable edge'' or ``stable boundary''.
\item $\Sigma=\sigma(-\frac12\Delta)$; this occurs only in $d_2\geq3$ and if
  $a$ is not too large, that is, if $a\in(0,a_0]$.\\
  In this case, both spectral edges are stable.
\end{enumerate}
On the other hand, it is well known (see~\cite{MR97c:47008}) that,
\begin{itemize}
\item if $d_2=1,2$, then, for $a>0$, $\sigma(-\frac12\Delta_{d_2}- a\Pi^2_0)
  =[-d_2,d_2]\cup\{\lambda (a)\}$, and the spectrum in $[-d_2,d_2]$ is
  purely absolutely continuous and $\lambda(a)$ is a simple
  eigenvalue;
\item if $d_2\geq3$, there exists $a^0>0$ such that
  \begin{itemize}
  \item if $0<a<a^0$, then,
    $\sigma(-\frac12\Delta_{d_2}-a\Pi^2_0)=[-d_2,d_2]$, and the spectrum is
    purely absolutely continuous;
  \item if $a=a^0$, then
    \begin{itemize}
    \item if $d_2=3,4$, then $\sigma(-\frac12\Delta_{d_2}-a\Pi^2_0)
      =[-d_2,d_2]$, the spectrum is purely absolutely continuous, and
      $-d_2$ is a resonance for $-\frac12\Delta_{d_2}-a\Pi^2_0$;
    \item if $d_2\geq5$, then $\sigma(-\frac12\Delta_{d_2}-a\Pi^2_0)
      =[-d_2,d_2]$, the spectrum is purely absolutely continuous in
      $[-d_2,d_2)$, and $-d_2$ is a simple eigenvalue for
      $-\frac12\Delta_{d_2}-a\Pi^2_0$;
    \end{itemize}
  \item if $a>a^0$, then,
    $\sigma(-\frac12\Delta_{d_2}-a\Pi^2_0)=[-d_2,d_2]\cup\{\lambda (a)\}$,
    and the spectrum in $[-d_2,d_2]$ is purely absolutely continuous
    and $\lambda(a)$ is a simple eigenvalue;
  \end{itemize}
\end{itemize}
For the operator $-\frac12\Delta_{d_2}+b\Pi^2_0$, we have a symmetric
situation.
\par Our aim is to study the density of surface states near the edges
of $\Sigma$. In the present case, both edges are obviously symmetric.
So we will only describe the lower edge. One has to distinguish
between the case of fluctuation and stable edges. The behavior in the
two cases are radically different.
\subsection{The stable edge}
\label{sec:stable-edge}
As the discussion for lower and upper edge are symmetric, let us
assume the lower edge is stable and work near that edge.\\
In the case of a stable edge, it is convenient to modify the
normalization of the IDSS. Therefore, we introduce the operator
\begin{equation*}
  H_t=-\frac12\Delta+t\car\otimes\Pi_0^2.
\end{equation*}
As above, let $a$ be the infimum of the random variables
$(\omega_j)_j$.  For $\varphi\in\Coi(\R)$, define
\begin{equation*}
  (\varphi,n_{s,\text{norm}})=
  \esp(\Tr(\Pi_1[\varphi(H_\omega)-\varphi(H_{a})]\Pi_1))
\end{equation*}
The advantage of this renormalization is that the IDSS
$n_{s,\text{norm}}$ is the distributional derivative of a positive
measure. Indeed, for $\varphi\in\Coi(\R)$, define
\begin{equation*}
  (\varphi,dN_{s,\text{norm}})=
  -\esp(\Tr(\Pi_1[P(\varphi)(H_\omega)-P(\varphi)(H_{a})]\Pi_1))
\end{equation*}
where
\begin{equation*}
  P(\varphi)(x)=\int_x^{+\infty}\varphi(t)dt.
\end{equation*}
Clearly, $dN_{s,\text{norm}}$ is independent of the anti-derivative of
$\varphi$ chosen to define it; it is a positive measure and
\begin{equation*}
  n_{s,\text{norm}}=-\frac{d}{dE}dN_{s,\text{norm}}.
\end{equation*}
Let $n_s^t$ be the IDSS for $H_t$. As above, one can define a
anti-derivative of $n_s^t$; denote it by $-dN_s^t$. Let
$n_{s,\text{norm}}^t$ be the normalized version of $n_s^t$, i.e.
$n_{s,\text{norm}}^t=n_s^t-n_s^{a}$. One has
\begin{equation}
  \label{eq:52}
  n_{s,\text{norm}}+n_s=n_s^{a}.
\end{equation}
One problem one encounters when studying $n_s$ is that very little is
known about its regularity for random surfacic models (see
nevertheless~\cite{MR1867350}). Thanks to~\eqref{eq:52}, we know that
$n_s$ is the difference of two distributions each of which is the
derivative of a signed measure. So we can take the counting function
of $dN_s$ as $dN_s=dN_{s,\text{norm}}-dN_s^{a}$ is the difference of
two measures. Thus, we define its counting function
\begin{equation}
  \label{eq:9}
  N_s(E)=\int_{-d}^E dN_s(e).
\end{equation}
An obvious consequence of~\eqref{eq:52} is the
\begin{Pro}
  \label{pro:3}
  One has
  \begin{equation}
     \label{eq:84}
     N_s^{a}(E)\leq N_s(E)\leq N_s^{b}(E).
  \end{equation}
\end{Pro}
\noindent In section~\ref{sec:asymptotics}, we study the asymptotics
for $N_s^t$. As a consequence of this study, we prove
\begin{Th}
  \label{thr:2}
  Assume $d_2=1$ or $2$. Then, one has
  \begin{equation*}
    N_s(E)\equ_{\substack{E\to -d\\E>-d}}
    \begin{cases}
      \D\frac{\text{Vol}(\mathbb{S}^{d_1-1})}
      {d_1(d_1+2)(2\pi)^{d_1}}\cdot(E+d)^{1+d_1/2} &\text{ if
      }d_2=1,\\
      \D\frac{2\text{Vol}(\mathbb{S}^{d_1-1})}{d_1(d_1+2
        )(2\pi)^{d_1}}\cdot \frac{(E+d)^{1+d_1/2}}{|\log(E+d)|}
      &\text{ if }d_2=2.
    \end{cases}
  \end{equation*}
  where $\mathbb{S}^{d_1-1}$ is the $d_1-1$ dimensional unit sphere.
\end{Th}
\noindent If $a>0$, this result is an immediate consequence of
Proposition~\ref{pro:3} and of Theorem~\ref{thr:3} giving the
asymptotics of the IDSS for constant surface potential (see also
section~\ref{sec:asymptotics}). If $a=0$, one needs to improve
upon~\eqref{eq:84} as the left hand side of this inequality vanishes
making it unusable. This is the purpose of Theorem~\ref{thr:1}.
\par When $d_2\geq3$, the situation becomes more complicated and we
are only able to use Proposition~\ref{pro:3} to get the two-sided
estimate
\begin{equation}
  \label{eq:85}
  C\frac{a(1+o(1))}{(1+aI)}
  \leq  \frac{(2\pi)^{d}}{s(E+d)(E+d)^{1+d_1/2}}\cdot N_s(E)
  \leq C\frac{b(1+o(1))}{(1+bI)}
\end{equation}
where $C$ is a positive constant depending only on the dimensions
$d_1$ and $d_2$ (see section~\ref{sec:asymptotics}) and
\begin{gather*}
  s(x)=\frac{1}{2}|x|^{\frac{d_2-2}{2}},\\
  I=\frac12\sup_{\theta_1\in\T^{d_1}}\int_{\theta_2\in\T^{d_2}}\left
    (d-\sum_{j=1}^{d_1}\cos(\theta^j_1)-
    \sum_{j=1}^{d_2}\cos(\theta_2^j)\right)^{-1}d\theta_2.
\end{gather*}
Here, and in the sequel, the measure $d\theta_{\alpha}$
($\alpha\in\{1,2\}$) is the Haar measure on the torus $\T^{d_\alpha}$,
i.e. the Lebesgue measure normalized to have total mass equal to one.
\par Let us note that, if $a<0<b$, the inequality~\eqref{eq:85} does
not give much information of the actual behavior of $N_s(E)$ when
$d_2\geq3$.
\subsection{The fluctuation edge}
\label{sec:fluctuation-edge}
Here, we assume that $E_0=\inf\sigma(H_\omega)$ is strictly below
$-d=\inf\sigma(-\frac12\Delta)$. In this case, $E_0$ is a fluctuation edge of
the spectrum.
\par Below the spectrum of $-\frac12\Delta$, the density of surface states
$n_s$ is positive; hence, it is a Borel measure and the integrated
density of surface states $N_s(E)$ can be defined as its distribution
function, i.e. $N_s(E)=n_s((-\infty,E))$ for $E<-d$. We will prove
Lifshitz type behavior for $N_s(E)$ for $E\searrow E_0$ which is
characteristic for fluctuation edges. However, the Lifshitz exponent,
in the homogeneous case typically equal to $-\frac{d}{2}$, is given by
$-\frac{d_1}{2}$ in our case. More precisely, we will show
\begin{equation*}
  \lim_{E\searrow E_0} \frac{\ln|\ln(N_s(E))|}{\ln(E-E_0)} =
  -\frac{d_1}{2}.
\end{equation*}

%
%% \tableofcontents
%
\section{The main results}
\label{sec:main-results}
Let us now describe the general model we consider. Let $H$ be a
translational invariant Jacobi matrix with exponential off-diagonal
decay that is $\D H=((h_{\gamma-\gamma'}))_{\gamma,\gamma'\in \Z^d}$
such that,
\begin{description}
\item[(H0.a)] $h_{-\gamma}=\overline{h_\gamma}$ for $\gamma\in\Z^d$
  and for some $\gamma\not=0$, $h_\gamma\not=0$.
\item[(H0.b)] There exists $c>0$ such that, for $\gamma\in\Z^d$,
    \begin{equation*}
      \vert h_\gamma\vert\leq \frac1ce^{-c\vert \gamma\vert}.
    \end{equation*}
\end{description}
The infinite matrix $H$ defines a bounded self-adjoint operator on
$\ell^2(\Z^d)$. Using the Fourier transform, it is easily seen that
$H$ is unitarily equivalent to the multiplication by the function
$\theta\mapsto h(\theta)$ defined by
\begin{equation*}
  h(\theta)=\sum_{\gamma\in\Z}h_\gamma e^{i\gamma\theta}\text{ where
  }\theta=(\theta_1,\dots,\theta_d),
\end{equation*}
acting as an operator on $L^2(\T^d)$ where $\T^d=\R^d/(2\pi\Z^d)$ (the
Lebesgue measure on $\T^d$ is normalized so that the constant function
1 has norm 1). The function $h$ is real analytic on $\T^d$. We
normalize it so that it be non-negative and $0$ be its minimum.\\
As both ends of the spectrum of our operator play symmetric parts, we
only study what happens at a left edge, i.e. near the bottom of the
spectrum. All our assumptions will reflect this fact.
\subsection{The case of a constant surface potential}
\label{sec:case-const-surf}
We will start with a study of the density of surface states when the
surfacic potential $V$ is constant, i.e. $V=t\Pi_0^2$. We define the
operator $H_t=H+t\car\otimes\Pi_0^2$. We prove two results on $H_t$.
The first one is a criterion for the positivity of $H_t$ and a
description of its infimum when it is negative; the other result
describes the density of the density of surface states near $0$ when
$H_t$ is non-negative.\\
In the present section, we assume
\begin{description}
\item[(H1)] the function $h:\ \T^d\to\R$ admits a unique minimum; it
  is quadratic non-degenerate.
\end{description}
If $H$ is $-\frac12\Delta$, then $h=h_0$ where
\begin{equation}
  \label{eq:7}
  h_0(\theta):=\cos(\theta_1)+\cdots+\cos(\theta_d).
\end{equation}
In this case, assumption (H1) is satisfied. Below, we give an example
why considering more general Hamiltonians can be of interest.\\
For the sake of definiteness, we assume the minimum of $h$ to be 0.
This amounts to adding a constant to $H$.
\par We start with a characterization of the infimum of the spectrum of
$H_t$. Therefore, write $h(\theta)=h(\theta_1,\theta_2)$ where
$\theta=(\theta_1,\theta_2)$, $\theta_1\in\T^{d_1}$,
$\theta_2\in\T^{d_2}$. Define
\begin{equation}
  \label{eq:25}
  I(\theta_1,z)=\int_{\T^{d_2}}\frac{1}{h(\theta_1,\theta_2)-z}d\theta_2.
\end{equation}
We recall that the measures $d\theta_2$ is normalized so that the
measure of $\T^{d_2}$ be equal to 1.\\
We prove
\begin{Pro}
  \label{pro:4}
  Assume (H0) and (H1) are satisfied.\\
  $H_t$ is non negative if and only if $t$ satisfies
  \begin{equation}
    \label{eq:19}
    1+tI_\infty\geq0\text{ where }I_\infty:=\sup_{\theta_1\in\T^{d_1}}
    \int_{\T^{d_2}}\frac{1}{h(\theta_1,\theta_2)}d\theta_2
  \end{equation}
  Assume now that $1+tI_\infty<0$. Then, there exists a unique
  $E_0\in(-\infty,0]$ such that
  \begin{equation*}
    \forall\theta_1\in\T^{d_1},\ 1+tI(\theta_1,E_0)\geq0\text{ and }
    \exists\theta_1\in\T^{d_1},\ 1+tI(\theta_1,E_0)=0.
  \end{equation*}
  Moreover, $E_0$ is the infimum of the spectrum of $H_t$.
\end{Pro}
\noindent Proposition~\ref{pro:4} is proved in
section~\ref{sec:dens-surf-stat}.
\par Criterion~\eqref{eq:19} immediately gives the obvious fact that
if $t\geq0$ then $H_t$ is non-negative. As we assumed that $h$ has
only non degenerate minima, if $d_2=1,2$ and $t<0$, then $H_t$ is not
non-negative.
\par We now turn to our second result. It describes the asymptotics of
$N_s^t$ near $0$ when~\eqref{eq:19} is satisfied. Recall that $N_s^t$
is the density of surface states of $H_t$.
\begin{Th}
  \label{thr:3}
  Assume $t$ satisfies condition~\eqref{eq:19}. Define
  \begin{equation*}
    I=\int_{\T^{d_2}}\frac{1}{h(0,\theta_2)}d\theta_2.
  \end{equation*}
  One has
  \begin{itemize}
  \item if $d_2=1$:
    \begin{equation*}
      \int_0^EdN_s^t(e)\equ_{E\to0^+}
      \D\frac{\text{Vol}(\mathbb{S}^{d_1-1})}
      {d_1(d_1+2)(2\pi)^{d_1}\sqrt{\text{Det}\,(Q_1-RQ_2^{-1}R^*)}} \cdot
      E^{1+d_1/2}
    \end{equation*}
  \item if $d_2=2$:
    \begin{equation*}
      \int_0^EdN_s^t(e)\equ_{E\to0^+} \frac{2\text{Vol}(\mathbb{S}^{d_1-1})}
      {d_1(d_1+2)(2\pi)^{d_1}\sqrt{\text{Det}\,(Q_1-RQ_2^{-1}R^*)}}
      \frac{E^{1+d_1/2}}{|\log E|}
    \end{equation*}
  \end{itemize}
  If $d_2\geq3$ and $1+t\cdot I>0$, then, one has
  \begin{equation*}
    \int_0^EdN_s^t(e)\equ_{E\to0^+}
    \frac{c(d_1,d_2)\text{Vol}(\mathbb{S}^{d_2-1})\text{Vol}(\mathbb{S}^{d_1-1})
    }{d(2\pi)^{d}\sqrt{\text{Det}\,Q}}\cdot \frac{t}{1+tI}\cdot s(E)E^{1+d_1/2}
  \end{equation*}
  If $d_2\geq3$ and $1+t\cdot I=0$, if we assume, moreover, that
  $\D\theta_1\mapsto I(\theta_1,0):=\int_{\T^{d_2}}
  (h(\theta_1,\theta_2))^{-1} d\theta_2$ has a local maximum for
  $\theta_1=0$, then one has
  \begin{itemize}
  \item if $d_2=3$:
    \begin{equation*}
      \int_0^EdN_s^t(e)de\equ_{E\to0^+}
      \D\frac{\D\int_{|\theta_1|\leq1}\text{Arg}(-i
        |1-\theta_1^2|^{1/2}+\tilde g(\theta_1))d\theta_1}
      {d_1(d_1+2)\pi(2\pi)^{d_1}\sqrt{\text{Det}\,(Q_1-RQ_2^{-1}R^*)}} \cdot
      E^{1+d_1/2}
    \end{equation*}
  \item if $d_2=4$:
    \begin{equation*}
      \int_0^EdN_s^t(e)\equ_{E\to0^+} -\frac{2\text{Vol}(\mathbb{S}^{d_1-1})}
      {d_1(d_1+2)(2\pi)^{d_1}\sqrt{\text{Det}\,(Q_1-RQ_2^{-1}R^*)}}
      \frac{E^{1+d_1/2}}{|\log E|}
    \end{equation*}
  \item if $d_2\geq5$:
    \begin{equation*}
      \int_0^EdN_s^t(e)\equ_{E\to0^+}
      \frac{c(d_1,d_2)\text{Vol}(\mathbb{S}^{d_2-1})\text{Vol}(\mathbb{S}^{d_1-1})
      }{d(2\pi)^{d}\sqrt{\text{Det}\,Q}}\cdot \frac{-1}{J}\cdot s(E)E^{d_1/2}
    \end{equation*}
  \end{itemize}
  Here, we used the following notations:
  \begin{itemize}
  \item Arg$(\cdot)$ denotes the principal determination of the
    argument of a complex number,
  \item for $n\in\{d_1,d_2\}$, $\mathbb{S}^{n-1}$ is the $n-1$
    dimensional unit sphere,
  \item $\tilde g$ is a linear form defined below,
  \item the function $s$ and the constants $c(d_1,d_2)$ and $J$ are
    defined by
    \begin{equation*}
      s(x)=\frac{1}{2}|x|^{\frac{d_2-2}{2}},\quad
      c(d_1,d_2)=\int_0^1r^{d_1-1}(1-r^2)^{(d_2-2)/2}dr,\quad
      J=\int_{\T^{d_2}}\frac{1}{h^2(0,\theta_2)}d\theta_1
    \end{equation*}
  \item $Q$ is the Hessian matrix of $h$ at $0$ that can be decomposed
    as $\D Q=\begin{pmatrix} Q_1 & R^*\\ R & Q_2 \end{pmatrix}$.
  \end{itemize}
\end{Th}
\noindent About the function $\tilde g$, it is defined as
follows. We assume $d_2\geq3$ and $1+tI=0$. Let $\D h_2(\theta_1)=
\inf_{\theta_2\in\T^{d_2}} h(\theta_1,\theta_2)$. In
section~\ref{sec:asymptotics}, we show that the function
$\D\theta_1\mapsto \int_{\T^{d_2}}(h(\theta_1,\theta_2)
-h_2(\theta_1))^{-1} d\theta_2$ is real analytic in a neighborhood of
$0$. Using the Taylor expansion of this function near $0$, one obtains
\begin{equation*}
  1+t\int_{\T^{d_2}}\frac{1}{h(\theta_1,\theta_2)-h_2(\theta_1)}
  d\theta_2= tg(\theta_1)+O(|\theta_1|^2).
\end{equation*}
This defines the linear form $g$ uniquely. Then, $\tilde g$ is defined
by
\begin{equation*}
  \tilde g(\theta):=(2\pi)^{d_2}\sqrt{\text{Det}\,(Q_2)}
  g((Q_1-RQ_2^{-1}R^*)^{-1/2}\theta_1).
\end{equation*}
If the variables $(\theta_1,\theta_2)$ separate in $h$, i.e., if
$h(\theta_1,\theta_2)=\tilde h_1(\theta_1)+\tilde h_2(\theta_2)$, the
function $\tilde g$ is identically 0.
\subsection{The case of a random surface potential}
\label{sec:case-rand-surf}
Let $V_\omega$ be a random potential concentrated on the sub-lattice
$\Z^{d_1}\times \{0\}\subset\Z^d$ ($d_1$ is chosen as in
section~\ref{sec:introduction}) of the form
\begin{equation}
  \label{eq:2}
  V_\omega(\gamma_1,\gamma_2)=\begin{cases}\omega_{\gamma_1}&\text{ if
   }\gamma_2=0,\\  0&\text{ if }\gamma_2\not=0.\end{cases},
   \gamma=(\gamma_1,\gamma_2)\in\Z^{d_1}\times \Z^{d_2}=\Z^d.
\end{equation}
and $(\omega_{\gamma_1})_{\gamma_1\in\Z^{d_1}}$ is a family of i.i.d.
bounded, non constant random variables.\\
Let $\omega_\pm$ be respectively the maximum and minimum of the random
variables $(\omega_{\gamma_1})_{\gamma_1\in\Z^{d_1}}$, and let
$\overline{\omega}$ be its expectation.
\par Finally, we define the random surfacic model by
\begin{equation}
  \label{eq:8}
  H_\omega=H+V_\omega,
\end{equation}
and its IDSS by
\begin{equation*}
  (n_s,\varphi)=\esp(\Tr(\Pi_1[\varphi(H_\omega)-\varphi(H)]\Pi_1))
\end{equation*}
Following section~\ref{sec:introduction}, one regularizes $n_s$ into
$N_s$ as in~\eqref{eq:9}.
\begin{Rem}
  \label{rem:1}
  An interesting case which can be brought back to a Hamiltonian of
  the form~\eqref{eq:8} with $H$ and $V_\omega$ as above is the
  following.\\
  Consider $\Gamma$, a sub-lattice of $\Z^d$ obtained in the following
  way $\Gamma=G(\{0\}\times\Z^{d_2})$ where $G$ is a matrix in
  $\text{GSL}_d(\Z)$, the $d$-dimensional special linear group over
  $\Z$, i.e. the multiplicative group of invertible matrices with
  coefficients in $\Z$ and unit determinant. One easily shows that the
  random operator
  \begin{equation*}
    H_\omega(\Gamma)=-\frac12\Delta
    +\sum_{\gamma\in\Gamma}\omega_\gamma\Pi_\gamma
  \end{equation*}
  (where $\Pi_\gamma$ is the projector onto the vector
  $\delta_\gamma\in\ell^2(\Z^d)$) is unitarily equivalent to
  $H+V_\omega$ where $V_\omega$ is defined in~\eqref{eq:2} and
  $h(\theta)=h_0(G'\cdot\theta)$; here, $h_0$ is defined
  in~\eqref{eq:7} and $G'$ is the inverse of the transpose of $G$,
  i.e. $G'=\,^tG^{-1}$.
\end{Rem}
\begin{Def}
  \label{def:1}
  We say that $E$, an edge (or boundary) of the spectrum of
  $H_\omega$, is {\it stable} if it is an edge of the spectrum of
  $H+tV_\omega$ for all $t\in[0,1]$. If an edge is not stable, we call
  it a {\it fluctuation} edge.
\end{Def}
\noindent Note that in the case of the introduction, this definition
is equivalent to that given there.\\
As in the introduction, one has to distinguish between
\begin{enumerate}
\item stable boundaries : at these boundaries, the IDSS is given by
  the IDSS of a model operator computed from the random model.
\item fluctuation boundaries: at these boundaries, one has standard
  Lifshitz tails.
\end{enumerate}
To complete this section, let us give a very simple description of the
spectrum of $H_\omega$. One has
\begin{Pro}
  \label{pro:2}
  Let $H_\omega$ be defined as above. Then
  \begin{equation*}
    \sigma(H_\omega)=\bigcup_{t\in\text{supp}(P_0)} \sigma(H_t).
  \end{equation*}
\end{Pro}
\noindent Here and in the following $P_0$ denotes the common
distribution of the random variables $(\omega_{\gamma_2})_{\gamma_2}$.
\subsection{The stable boundaries}
\label{sec:non-fluct-bound}
The stable boundary we are studying is the lower boundary that we
assumed to be 0. Let us first give a criterion for the lower edge of
the spectrum of $H$ (that we assume to be equal to 0) to be a stable
edge. We prove
\begin{Pro}
  \label{pro:1}
  Write $h(\theta)=h(\theta_1,\theta_2)$ where
  $\theta=(\theta_1,\theta_2)$, $\theta_1\in\T^{d_1}$,
  $\theta_2\in\T^{d_2}$. Then, $0$ is a stable spectral edge if and
  only if $\omega_-$ satisfies condition~\eqref{eq:19}.
\end{Pro}
\noindent Proposition~\ref{pro:1} is an immediate consequence of
Proposition~\ref{pro:4} and Proposition~\ref{pro:2}. It gives the
obvious fact that, if $\omega_-\geq0$, then $0$ is a stable edge. As
we assumed that $h$ has only non degenerate minima, we see that if
$d_2=1,2$ and $\omega_-<0$, then $0$ is never a stable edge.
Actually, it need not be an edge of the spectrum of $H_\omega$.
\par Using the same notations as above, we prove
\begin{Th}
  \label{thr:1}
  Assume (H0) and (H1) are verified.  Assume, moreover, that $0$ is a
  stable spectral edge for $H_\omega$. Then, one has
    \begin{equation}
      \label{eq:75}
      \text{if }\overline{\omega}>0,\text{ then }
      \liminf_{E\to0^+}\frac{N_s(E)}{N_s^{\overline{\omega}}(E)}\geq 1
      \quad\text{and\quad if }\overline{\omega}<0,\text{ then }
      \limsup_{E\to0^+}\frac{N_s(E)}{N_s^{\overline{\omega}}(E)}\leq 1
    \end{equation}
    where $N_s^{\overline{\omega}}$ is the IDSS of the operator with
    constant surface potential $\overline{\omega}$, the common
    expectation value of the random variables
    $(\omega_{\gamma_1})_{\gamma_1}$.
\end{Th}
\noindent This result admits an immediate corollary
\begin{Th}
  \label{thr:4}
  Assume (H0) and (H1) hold.  Assume, moreover, that $0$ is a
  stable spectral edge for $H_\omega$. Then,
  \begin{itemize}
  \item if $d_2=1$:
    \begin{equation*}
      N_s(E)\equ_{E\to0^+}
      \D\frac{\text{Vol}(\mathbb{S}^{d_1-1})}
      {d_1(d_1+2)(2\pi)^{d_1}\sqrt{\text{Det}\,(Q_1-RQ_2^{-1}R^*)}} \cdot
      E^{1+d_1/2};
    \end{equation*}
  \item if $d_2=2$:
    \begin{equation*}
      N_s(E)\equ_{E\to0^+} \frac{2\text{Vol}(\mathbb{S}^{d_1-1})}
      {d_1(d_1+2)(2\pi)^{d_1}\sqrt{\text{Det}\,(Q_1-RQ_2^{-1}R^*)}}
      \frac{E^{1+d_1/2}}{|\log E|}.
    \end{equation*}
  \end{itemize}
\end{Th}
\noindent Theorem~\ref{thr:4} is an immediate consequence of
Theorem~\ref{thr:1} and the bound
\begin{equation*}
  N_s^{\omega_-}(E)\leq N_s(E)\leq N_s^{\omega_+}(E).
\end{equation*}
As noted in the introduction, Theorem~\ref{thr:1} is only necessary
when $\omega_-=0$ (in which case $\overline{\omega}>0$). Moreover, one
obtains the analogue of~\eqref{eq:85} in the present case for $d_2\geq
3$.
\par The above results may lead to the belief that
\begin{equation*}
  N_s(E)\equ_{E\to0} N_s^{\overline{\omega}}(E)
\end{equation*}
for all dimensions $d_2$. Let us now explain why this result, if true,
is not obtained for dimension $d_2\geq3$.  Therefore, we explain the
heuristics behind the proof of Theorem~\ref{thr:1}; it is very similar
to that of standard Lifshitz tails with one big difference when
$d_2\geq3$.\\
Restrict $H_\omega$ to some large cube. One wants to estimate the IDSS
for $H_\omega$; for this restriction, this comes up to estimating the
differences between the integrated density of states (the usual one)
of the operator $H_\omega$ and the integrated density of the operator
$H_{\omega_-}$ (see Lemma~\ref{estapp}). So we want to count the
eigenvalues of $H_\omega$ below energy $E$, say, subtract the number
of eigenvalues of $H_{\omega_-}$ below energy $E$, divide by the
volume of the cube, and see how this behaves when $E$ gets small.
Assume $\varphi$ is a normalized eigenfunction associated to an
eigenvalue of $H_\omega$ below $E$. Then, one has
$\langle(H+V_\omega)\varphi,\varphi\rangle\geq E$. Assume for a moment
that $V_\omega$ is non negative. Then, we see that one must have both
$\langle H\varphi,\varphi\rangle\geq E$ and $\langle
V_\omega\varphi,\varphi\rangle\geq E$. The first of these conditions
guarantees that $\varphi$ is localized in momentum. So it has to be
extended in space. If one plugs this information into the second
condition, one sees that $\langle V_\omega\varphi,\varphi\rangle
\sim\overline{\omega}$ with a large probability. So that, to
$\varphi$, $H_\omega$ roughly looks like $H+\overline{\omega}\Pi_0^2$.
There is one problem with this reasoning:\\
as $V_\omega$ only lives on a hyper-surface, and as $\varphi$ is flat,
it only sees a very small part of $\varphi$; a simple calculation
shows that $\Vert\Pi_ 0^2\varphi\Vert\sim E^{d_2/2}$; on the other
hand, when one says that $\varphi$ is roughly constant, one makes an
error of size $E^\alpha$ (for some $0<\alpha<1$); hence, for dimension
$d_2\geq3$, this error is much larger than the term we want to
estimate, namely, $\langle V_\omega\varphi,\varphi\rangle$. In other
words, because $\varphi$ is very flat, we can modify it on the
hyper-surface (e.g. localize the part of it living on the
hyper-surface) with almost no change to the total energy of $\varphi$;
hence, we cannot guarantee that $\varphi$ is also flat on the
hyper-surface, which implies that $\langle
V_\omega\varphi,\varphi\rangle$ need not be close $\overline{\omega}$
with a large probability.
\subsection{The fluctuation boundaries}
\label{sec:fluct-bound-that}
In this section we assume that the infimum of $\Sigma$ which we call
$E_0$ is (strictly) below $\inf(\sigma(H))$, so that $E_0$ is a
fluctuation edge.  In this case, we consider a ``reduced'' operator
$\tilde{H}$ which acts on $\ell^2(\Z^{d_1})$. In Fourier
representation this operator is multiplication by the function
$\tilde{h}$ given by:
\begin{equation}
  \label{eq:3}
  \tilde{h}(\theta_1)=\left(\int_{\T^{d_2}}
    \frac{1}{h(\theta_1,\theta_2)-E_0}\;
  d\theta_2\right)^{-1} + E_0
\end{equation}
We will reduce the proof of Lifshitz tails for $H_\omega=H+V_\omega$
to a proof of Lifshitz tails for the reduced operator
$\tilde{H}_\omega = \tilde{H} +\tilde V_\omega$ (where $\tilde
V_\omega$ is a diagonal matrix with entries
$(\omega_{\gamma_1})_{\gamma_1}$). To prove Lifshitz tail behavior for
$\tilde{H}_\omega$ we have to impose a condition on the behavior of
$\tilde{h}$ near its minimum. We either suppose:
\begin{description}
\item[(H2)] the function $\tilde{h}:\ \T^{d_1}\to\R$ admits a unique
  quadratic minimum.
\end{description}
or we assume the weaker hypothesis:
\begin{description}
\item[(H2')] the function $\tilde{h}:\ \T^d\to\R$ is not constant.
\end{description}
Moreover, we always assume that the random variables
$\omega_{\gamma_1}$ defining the potential (\ref{eq:V}) are
independent with a common distribution $P_0$. We set
$\omega_-=\inf(\text{supp}(P_0))$ and assume:
\begin{description}
\item[(H3)] $P_0$ is not concentrated in a single point and
  $P_0([\omega_-,\omega_-+\varepsilon))\ge C\; \varepsilon^k$ for some
  $k$.
\end{description}
We will prove below:
\begin{Th}
  \label{thr:8}
  If (H2) and (H3) are satisfied then
  \begin{equation*}
    \lim_{E\searrow E_0}\frac{\ln|\ln(N_s(E))|}{\ln(E-E_0)} =
    -\frac{d_1}{2}.
  \end{equation*}
\end{Th}
\noindent We have an additional result for low dimension of the surface:
\begin{Th}
  \label{thr:9}
  Assume (H2') and (H3) hold. If $d_1=1$ then
  \begin{equation*}
    \lim_{E\searrow E_0} \frac{\ln |\ln(N_s(E))|}{\ln(E-E_0)} =
    -\lim_{E\searrow E_0}\frac{\ln(n(E))}{(E-E_0)}
  \end{equation*}
  where $n(E)$ is the integrated density of states for $\tilde{H}$.\\
  If $d_2=2$, then
  \begin{equation*}
    \lim_{E\searrow E_0} \frac{\ln |\ln(N_s(E))|}{\ln(E-E_0)} =
    -\alpha
  \end{equation*}
  where the computation of $\alpha$ is explained below.
\end{Th}
\noindent For the sake of simplicity, let us assume $E_0=0$. The
Lifshitz exponent $\alpha$ will depend on the way $\tilde{h}$ vanishes
at $\mathcal{S}=\{\theta_1 | \tilde{h} = 0 \}$ and on the curvature of $\mathcal{S}$.
\par To describe it precisely, we need to introduce some objects from
analytic geometry (see~\cite{MR1979772} for more details). If
$\mathcal{E}$ is a set contained in the closed first quadrant in
$\R^2$ then its {\it exterior convex hull} is the convex hull of the
union of the rectangles $R_{xy}=[x,\infty)\times  [y,\infty)$, where the
union is taken over all $(x,y)\in \mathcal{E}$.
\par Pick $\theta_0\in\mathcal{S}$ and consider the Newton diagram of
$\tilde{h}$ at $\theta_0$, i.e.,
\begin{enumerate}
\item Express $\tilde{h}$ as a Taylor series at $\theta_0$,
  $\tilde{h}(\theta^1,\theta^2)=\sum_{ij}a_{ij}(\theta^1-\theta_0^1)^i
  (\theta^2-\theta_0^2)^j$, $\theta=(\theta^1,\theta^2)$.
\item Form the exterior convex hull of the points $(i,j)$ with
  $a_{ij}\neq 0$. This is a convex polygon, called the {\it Newton
    polygon}.
\item The boundary of the polygon is the {\it Newton diagram}.
\end{enumerate}
The Newton decay exponent is then defined as follows. The Newton
diagram consists of certain line segments. Extend each to a complete
line and intersect it with the diagonal line $\theta^1=\theta^2$.
This gives a collection of points $(a_k,a_k)$, one for each boundary
segment. Take the reciprocal of the largest $a_k$ and call this number
$\tilde{\alpha}(\tilde{h},\theta_0)$; it is the {\it Newton decay
  exponent}. Define
$\alpha(\tilde{h},\theta_0)=\min\{\tilde{\alpha}(\tilde{h}\circ T_0,
\theta_0): \, T_0(\cdot )=\theta_0+T(\cdot -\theta_0),\ T\in SL(2,\R)\}$.
\par Similarly, define $\alpha(\tilde{h},\theta)$ if $\theta$ is any
other point in $\mathcal{S}$, the zero set of $\tilde{h}$. Then, the
{\it Lifshitz exponent} $\alpha$ is defined by
\begin{equation}
  \label{eq:86}
  \alpha=\min_{\theta\in\mathcal{S}}\alpha(\tilde{h},\theta).
\end{equation}
The Lifshitz exponent $\alpha$ is positive as $\theta\mapsto
\alpha(\tilde{h},\theta)$ is a positive, lower semi-continuous
function and $\mathcal{S}$ is compact (see~\cite{MR1979772}).
\begin{Rem}
  \label{rem:4}
  Let us return to the example given in Remark~\ref{rem:1}. In the
  section~\ref{sec:appendix}, we check that (H.2') holds in this case;
  so for $d=d_1+d_2=3$, Theorem~\ref{thr:9} applies.
\end{Rem}
\section{Approximating the IDSS}
\label{sec:approximating-dss}
To approximate the IDSS, we use a method that has proved useful to
approximate the density of states of random Schr{\"o}dinger operators, the
periodic approximations. We shall show that the IDSS is well
approximated by the suitably normalized density of states of a well
chosen periodic operator.
\subsection{Periodic approximations}
\label{sec:peri-appr}
Let $(\omega_{\gamma_1})_{\gamma_1\in\Z^{d_1}}$ be a realization of
the random variables defined above. Fix $N\in\N^*$. We define
$H_{\omega}^N$, a periodic operator acting on $\ell^2(\Z^d)$ by
\begin{equation*}
  H_{\omega}^N=H+V_{\omega}^N=H+\sum_{\gamma_1\in\Z^{d_1}_{2N+1}}\omega_n
  \sum_{\substack{\beta_1\in(2N+1)\Z^{d_1}\\\beta_2\in(2N+1)\Z^{d_2}}}
  \vert\delta_{\gamma_1+\beta_1}\otimes\delta_{\beta_2}
  \rangle\langle\delta_{\gamma_1+\beta_1}\otimes\delta_{\beta_2}\vert.
\end{equation*}
Here, $\Z^{\tilde d}_{2N+1}=\Z^{\tilde d}/(2N+1)\Z^{\tilde d}$,
$\D\delta_l= (\delta_{jl})_{j\in\Z^{\tilde d}}$ is a vector in the
canonical basis of $\ell^2(\Z^{\tilde d})$ where $\delta_{jl}$ is the
Kronecker symbol and, $\tilde d=d_1$ or $\tilde d=d_2$, the choice
being clear from the context. As usual, $\vert u\rangle\langle u\vert$
is the orthogonal projection on a unit vector $u$.\\
By definition, $H_{\omega}^N$ is periodic with respect to the (non
degenerate) lattice $(2N+1)\Z^d$. We define the density of states
denoted by $n^N_{\omega}$ as usual for periodic operators: for
$\varphi\in\Coi(\R)$,
\begin{equation*}
  (\varphi,dn^N_{\omega})=\int_\R\varphi(x)dn^N_{\omega}(x)
  =\lim_{L\to+\infty}\frac1{(2L+1)^{d}}\sum_{\substack{\gamma\in\Z^{d}
        \\|\gamma|\leq L}}\langle\delta_{\gamma},\varphi(H_{\omega}^N)
    \delta_{\gamma}\rangle.
\end{equation*}
This limit exists (see e.g.~\cite{Cy-Fr-Ki-Si:87,MR94h:47068}). In a
similar way, one can define the density of states of $H$; we denote it
by $dn_0$. The operators $(H_\omega^N)_{\omega,N}$ are uniformly
bounded; hence, their spectra are contained in a fixed compact set,
say $\mathcal{C}$.  This set also contains the spectrum of $H_\omega$
and $H$. We prove
\begin{Le}
  \label{le:3}
  Pick $\mathcal{U}\subset\R$ a relatively compact open set such that
  $\mathcal{C}\subset\mathcal{U}$. There exists $C>1$ such that, for
  $\varphi\in\Coi(\R)$, for $K\in\N$, $K\geq1$, and $N\in\N^*$, we
  have
  \begin{equation}\label{appper}
    \left\vert(\varphi,dn)-(2N+1)^{d_2}\esp\{(\varphi,[dn^N_{\omega}-dn_0])\}
      \right\vert\leq
      \left(\frac{CK}N\right)^K\sup_{\substack{x\in\mathcal{U}\\0\leq
      J\leq K+d+2}}\left\vert\frac{d^J\varphi}{d^Jx}(x)\right\vert.
  \end{equation}
\end{Le}
\noindent{\bf Proof of Lemma~\ref{le:3}}  Fix $\varphi\in\Coi(\R)$. As
the spectra of the operators $H_\omega^N$ are contained in
$\mathcal{U}$, we may restrict ourselves to $\varphi$ supported in
$\mathcal{U}$ which we do from now on. By the
definition~\eqref{eq:11}, one has
\begin{equation}
  \label{eq:16}
    (\varphi,n_s)=\esp\left(\sum_{\gamma\in\Z^{d_2}}
      \langle\delta_0\otimes\delta_{\gamma_2},[\varphi(H_\omega)-
      \varphi(H)]\delta_0\otimes\delta_{\gamma_2}\rangle\right)
    =M_N(\varphi)+R_N(\varphi)
\end{equation}
where
\begin{gather*}
  M_N(\varphi)=\esp\left(\sum_{\substack{\gamma_2\in\Z^{d_2}\\
        \vert\gamma_2\vert\leq N}}
    \langle\delta_0\otimes\delta_{\gamma_2},[\varphi(H_\omega)-
    \varphi(H)]\delta_0\otimes\delta_{\gamma_2}\rangle\right),
  \\
  R_N(\varphi)=\esp\left(\sum_{
      \substack{\gamma_2\in\Z^{d_2}\\\vert\gamma_2\vert>N}}
    \langle\delta_0\otimes\delta_{\gamma_2},[\varphi(H_\omega)-
    \varphi(H)]\delta_0\otimes\delta_{\gamma_2}\rangle\right).
\end{gather*}
Let us now show that
\begin{equation}
  \label{eq:12}
  |R_N(\varphi)|\leq\left(\frac{CK}N\right)^K
  \sup_{\substack{x\in\mathcal{U}\\0\leq J\leq
      K+d+2}}\left\vert\frac{d^J\varphi}{d^Jx}(x)\right\vert.
\end{equation}
Therefore, we use some ideas from the proof of Lemma 1.1
in~\cite{Kl:97a}. Helffer-Sj{\"o}strand's formula (\cite{He-Sj:90}) reads
\begin{equation*}
  \varphi(H_{\omega})=
  \frac i{2\pi}\int_\C \frac{\partial\tilde\varphi}{\partial{\overline
  z}}(z)\cdot  (z-H_{\omega})^{-1}dz\wedge d{\overline z}.
\end{equation*}
where $\tilde\varphi$ is an almost analytic extension of $\varphi$
(see~\cite{Mat:71}), i.e. a function satisfying
\begin{enumerate}
\item for $z\in\R$, $\tilde\varphi(z)=\varphi(z)$;
\item supp$(\tilde\varphi)\subset\{z\in\C;\
  \vert\text{Im}(z)\vert<1\}$;
\item $\tilde\varphi\in{\mathcal S}(\{z\in\C;\
  \vert\text{Im}(z)\vert<1\})$;
\item the family of functions $\D
  x\mapsto\frac{\partial\tilde\varphi}{\partial\overline z}(x+iy)\cdot
  \vert y\vert^{-n}$ (for $0<\vert y\vert<1$) is bounded in $\Sc(\R)$
  for any $n\in{\mathbb N}$; more precisely, there exists $C>1$ such
  that, for all $p,q,r\in\N$, there exists $C_{p,q}>0$ such that
  \begin{equation}
    \label{eq:43}
    \sup_{0<\vert y\vert\leq 1}\sup_{x\in\R}\left\vert
      x^p\frac{\partial^q}{\partial x^q}\left(\vert
        y\vert^{-r}\cdot  \frac{\partial\tilde\varphi}{\partial\overline
          z}(x+iy)\right)\right\vert\leq
    C^r C_{p,q}\sup_{\substack{q'\leq r+q+2 \\
        p'\leq p}}\sup_{x\in\R}\left\vert
      x^{p'}\frac{\partial^{q'}\varphi}{\partial
        x^{q'}}(x)\right\vert.
  \end{equation}
\end{enumerate}
As we are working with $\varphi$ with compact support in
$\mathcal{U}$, its almost analytic extension can be taken to have
support in $(\mathcal{U}+[-1,1])+i[-1,1]$ (see
e.g.~\cite{Di-Sj:98}).\\
We estimate $\esp(|\langle\delta_0\otimes\delta_{\gamma_2},
[\varphi(H_\omega)-\varphi(H)]\delta_0\otimes\delta_{\gamma_2}\rangle|)$
for $\vert\gamma_2\vert>N$. Using the fact that the random variables
$(\omega_{\gamma_2})_{\gamma_2}$ are bounded, we get
\begin{equation*}
  \begin{split}
    &\esp(|\langle\delta_0\otimes\delta_{\gamma_2},[\varphi(H_\omega)-
    \varphi(H)]\delta_0\otimes\delta_{\gamma_2}\rangle|)\\&\leq
    \frac1{4\pi}\esp\left(\int_{\C}\left\vert\frac{\partial\tilde\varphi}
        {\partial{\overline z}}(z)\right\vert
      \vert\langle\delta_0\otimes\delta_{\gamma_2},\left((z-H_{\omega}^N)^{-1}-
        (z-H)^{-1}\right)\delta_0\otimes\delta_{\gamma_2}\rangle\vert
      dxdy\right) \\&\leq C\sum_{\gamma_1\in\Z^{d_1}}
    \int_{\C}\left\vert\frac{\partial\tilde\varphi}
      {\partial{\overline z}}(z)\right\vert\cdot\esp\left(
      \vert\langle\delta_0\otimes\delta_{\gamma_2},(z-H_{\omega}^N)^{-1}
      \delta_{\gamma_1}\otimes\delta_0\rangle\vert\cdot \vert\langle\delta_{\gamma_1}
      \otimes\delta_0, (z-H)^{-1}\delta_0\otimes\delta_{\gamma_2}
      \rangle\vert\right) dxdy
  \end{split}
\end{equation*}
where $z=x+iy$.\\
By a Combes-Thomas argument (see e.g.~\cite{Kl:01a}), we know that
there exists $C>1$ such that, uniformly in $(\omega_{\gamma})_\gamma$,
$\gamma_1\in\Z^{d_1}$ and $N\geq1$, we have, for Im$(z)\neq0$,
\begin{multline}
  \label{eq:18}
  \vert\langle\delta_{\gamma_1}\otimes\delta_{\gamma_2},
  (z-H_{\omega}^N)^{-1}\delta_{\gamma'_1}\otimes\delta_{\gamma'_2}\rangle\vert+
  \vert\langle\delta_{\gamma_1}\otimes\delta_{\gamma_2},
  (z-H)^{-1}\delta_{\gamma'_1}\otimes\delta_{\gamma'_2}\rangle\vert\leq\\
  \leq \frac C{\vert\text{Im}(z)\vert}
  e^{-\vert\text{Im}(z)\vert(|\gamma_1-\gamma'_1|+|\gamma_2-\gamma'_2|)/C}
\end{multline}
Hence, for some $C>1$,
\begin{equation*}
  \begin{split}
    |R_N(\varphi)|&\leq C\sum_{\gamma_1\in\Z^{d_1}}
    \int_{\C}\left\vert\frac{\partial\tilde\varphi}
      {\partial{\overline z}}(z)\right\vert\cdot
    \frac1{\vert\text{Im}(z)\vert^2}
    e^{-\vert\text{Im}(z)(|\gamma_1|+|\gamma_2|)\vert/C}
    dxdy\\
    &\leq C \int_{\C}\left\vert\frac{\partial\tilde\varphi}
      {\partial{\overline z}}(z)\right\vert
    \frac1{\vert\text{Im}(z)\vert^{d+2}}
    e^{-\vert\text{Im}(z)N\vert/C}dxdy.
  \end{split}
\end{equation*}
Taking into account the properties of almost analytic
extensions~\eqref{eq:43}, for some $C>1$, for $K\geq1$ and $N\geq1$,
we get
\begin{equation*}
  \begin{split}
    |R_N(\varphi)|&\leq C^{K+1}
    \int_{(\mathcal{U}+[-1,1])+i[-1,1]}\vert y\vert^Ke^{-\vert yN
      \vert/C}dxdy\sup_{\substack{x\in\mathcal{U}\\0\leq J\leq
        K+d+2}}\left\vert\frac{d^J\varphi}{d^Jx}(x)\right\vert
    \\&\leq\left(\frac{CK}{N}\right)^K\sup_{\substack{x\in\mathcal{U}\\0\leq
        J\leq K+d+2}}\left\vert\frac{d^J\varphi}{d^Jx}(x)\right\vert.
  \end{split}
\end{equation*}
This completes the proof of~\eqref{eq:12}.%
\par We now compare $M_N(\varphi)$ to
$(2N+1)^{d_2}\esp\{(\varphi,[dn^N_{\omega}-dn_0])\}$. Therefore, we
rewrite this last term as follows. Using the $(2N+1)\Z^d$ periodicity
of $H_\omega^N$ and $H$, we get
\begin{equation*}
    \sum_{\substack{\gamma\in\Z^{d}\\|\gamma|\leq N+L(2N+1)}}
  \langle\delta_{\gamma},\varphi(H_{\omega}^N)
    \delta_{\gamma}\rangle=(2L+1)^d
    \sum_{\substack{\gamma\in\Z^{d}\\|\gamma|\leq N}}
  \langle\delta_{\gamma},\varphi(H_{\omega}^N)
    \delta_{\gamma}\rangle.
\end{equation*}
This gives
\begin{equation}
  \label{eq:13}
  (2N+1)^d(\varphi,dn^N_\omega)=\esp\left(
    \sum_{\substack{\gamma\in\Z^{d}\\|\gamma|\leq N}}
  \langle\delta_{\gamma},\varphi(H_{\omega}^N)
    \delta_{\gamma}\rangle\right).
\end{equation}
On the other hand, as the random variables
$(\omega_{\gamma_2})_{\gamma_2}$ are i.i.d. and as $H$ is
$\Z^d$-periodic, as in~\cite{Kl:01a}, one computes
\begin{equation*}
  \begin{split}
    \esp\left( \sum_{\substack{\gamma\in\Z^{d}\\|\gamma|\leq N}}
      \langle\delta_{\gamma},\varphi(H_{\omega}^N)
      \delta_{\gamma}\rangle\right)&= \esp\left(\sum_{
        \substack{\gamma_1\in\Z^{d_1},\ \vert\gamma_1\vert\leq N
          \\\gamma_2\in\Z^{d_2},\ \vert\gamma_2\vert\leq N}}
      \langle\delta_{\gamma_1}\otimes\delta_{\gamma_2},\varphi(H_{\omega}^N)
      \delta_{\gamma_1}\otimes\delta_{\gamma_2}\rangle\right)\\&
    =(2N+1)^{d_1}\esp\left(\sum_{ \substack{\gamma_2\in\Z^{d_2}\\
          \vert\gamma_2\vert\leq N}}
      \langle\delta_0\otimes\delta_{\gamma_2},\varphi(H_{\omega}^N)
      \delta_0\otimes\delta_{\gamma_2}\rangle\right)
  \end{split}
\end{equation*}
Combining this with~\eqref{eq:13}, we get
\begin{equation*}
  (2N+1)^{d_2}\esp[(\varphi,dn^N_\omega)]=
  \esp\left(\sum_{ \substack{\gamma_2\in\Z^{d_2}\\
          \vert\gamma_2\vert\leq N}}
      \langle\delta_0\otimes\delta_{\gamma_2},\varphi(H_{\omega}^N)
      \delta_0\otimes\delta_{\gamma_2}\rangle\right)
\end{equation*}
Of course, such a formula also holds when $H_\omega^N$ is replaced
with $H$. In view of~\eqref{eq:11},~\eqref{eq:12} and~\eqref{eq:16},
to complete the proof of Lemma~\ref{le:3}, we need only to prove
\begin{equation}
  \label{eq:17}
  \esp\left\vert\sum_{ \substack{\gamma_2\in\Z^{d_2}\\
        \vert\gamma_2\vert\leq
        N}}\langle\delta_0\otimes\delta_{\gamma_2},
      [\varphi(H_{\omega}^N)-\varphi(H_{\omega})]
      \delta_0\otimes\delta_{\gamma_2}\rangle\right\vert\leq
    \left(\frac{CK}N\right)^K
    \sup_{\substack{x\in\mathcal{U}\\0\leq J\leq
        K+d+2}}\left\vert\frac{d^J\varphi}{d^Jx}(x)\right\vert.
\end{equation}
for $\varphi$, $K$ $J$ and $N$ as in Lemma~\ref{le:3}.\\
Proceeding as above, for $\gamma_2\in\Z^{d_2}$,
$\vert\gamma_2\vert\leq N$, we estimate
\begin{multline*}
  |\langle\delta_0\otimes\delta_{\gamma_2},
  [\varphi(H_{\omega}^N)-\varphi(H_{\omega})]
  \delta_0\otimes\delta_{\gamma_2}\rangle|
  \\\leq C\left[\sum_{\substack{\gamma'_1\in\Z^{d_1}\\
        \gamma'_2\in((2N+1)\Z^{d_2})^*}}
    +\sum_{\substack{\gamma'_1\in\Z^{d_1},\
        |\gamma'_1|>N\\\gamma'_2=0}} \right]
  \int_{\C}\left\vert\frac{\partial\tilde\varphi} {\partial{\overline
        z}}(z)\right\vert dxdy\cdot
  \\\esp\left(\vert\langle\delta_0\otimes\delta_{\gamma_2},(z-H_{\omega}^N)^{-1}
    \delta_{\gamma'_1}\otimes\delta_{\gamma'_2}\rangle\vert\cdot\right.\\
  \left.\vert\langle\delta_{\gamma'_1}\otimes\delta_{\gamma'_2},
    (z-H_\omega)^{-1}\delta_0\otimes\delta_{\gamma_2}\rangle\vert\right).
\end{multline*}
Here we used the fact that the operators $H_\omega$ and $H_\omega^N$
coincide in the cube $\{|\gamma|\leq N\}$.\\
As $H_\omega$ satisfies the same Combes-Thomas estimate~\eqref{eq:18}
as $H_\omega^N$, doing the same computations as in the estimate for
$R_N(\varphi)$, we obtain~\eqref{eq:17}. This completes the proof of
Lemma~\ref{le:3}.\qed
\vskip.2cm\noindent Obviously, one has an analogue of~\eqref{appper}
for $n_{s,\text{norm}}$, $n^t_s$ or $n^t_{s,\text{norm}}$. One needs
to replace $H_\omega^N$ and $H$ with their obvious counterparts i.e.,
choose the random variables $(\omega_{\gamma_2})_{\gamma_2}$ to be the
appropriate constant.
\par This enables us to prove
\begin{Le}
  \label{estapp}
  Fix $I$, a compact interval. Pick $\alpha>0$. There exists $\nu_0>0$
  and $\rho>0$ such that, for $\gamma\in[0,1]$, $E\in I$, $\nu\in(0,
  \nu_0)$ and $N\geq \nu^{-\rho}$, one has
  \begin{equation}
    \label{estpre}
      \esp(N^N_{\omega}(E-\nu))- e^{-\nu^{-\alpha}}\leq
      N_s(E)\leq
      \esp(N^N_{\omega}(E+\nu))+e^{-\nu^{-\alpha}}
  \end{equation}
  where $N^N_{\text{norm},\omega}=N^N_\omega-N^N_{\omega_-}$, and
  $N^N_\omega$ (resp. $N^N_{\omega_-}$) is the integrated density of
  states of $H_\omega^N$ (resp $H^N_{\omega_-}$, i.e. $H_\omega^N$
  where $\omega_\gamma=\omega_-$ for all $\gamma$).
\end{Le}
\noindent Let us note here that one can prove a similar result for the
approximation of $N_{s,\text{norm}}$ by $N^N_{\text{norm},\omega}$ or
that of $N_s^t$ by $N_s^{t,N}$.\\
{\bf Proof} Let us now prove Lemma~\ref{estapp}. Pick $\varphi$ a
Gevrey class function of Gevrey exponent $\alpha>1$
(see~\cite{Hor:83}); assume, moreover, that $\varphi$ has support in
$(-1,1)$, that $0\leq\varphi\leq1$ and that $\varphi\equiv1$ on
$(-1/2,1/2]$. Let $E\in I$ and $\nu\in(0,1)$,and set
\begin{equation*}
  \varphi_{E,\nu}(\cdot  )=\car_{[0,E]}*\varphi\left(\frac\cdot  \nu\right).
\end{equation*}
Then, by Lemma~\ref{le:3} and the Gevrey estimates on the derivatives
of $\varphi$, there exist $C>1$ such that, for $N\geq1$, $k\geq1$ and
$0<\nu<1$, we have
\begin{equation}
  \label{eq:10}
  \vert\esp((\varphi_{E,\nu},dN^N_{\omega}))-
  (\varphi_{E,\nu},dN_s)\vert\leq
  C(N\nu)^{3}\left(\frac{Ck^{1+\alpha}}{N\nu}\right)^k.
\end{equation}
We optimize the right hand side of~\eqref{eq:10} in $k$ and get that,
there exist $C>1$ such that, for $N\geq1$ and $0<\nu<1$, we have
\begin{equation*}
  \vert\esp((\varphi_{E,\nu},dN^N_{\omega}))-
  (\varphi_{E,\nu},dN_s)\vert\leq
  C(N+\nu^{-1})^3 e^{-(N\nu/C)^{1/(1+\alpha)}+
    C(N\nu/C)^{-1/(1+\alpha)}}
\end{equation*}
Now, there exist $\nu_0>0$ such that, for $0<\nu<\nu_0$ and $\D
N\geq\nu^{-1-\eta}$, we have
\begin{equation}
  \label{eq:22}
  \vert\esp((\varphi_{E,\nu},dN^N_{\omega}))-
  (\varphi_{E,\nu},dN_s)\vert\leq
   e^{-\nu^{-\eta/(2\alpha)}}.
\end{equation}
By definition, $\varphi_{E,\nu}\equiv1$ on $[0,E]$, and
$\varphi_{E,\nu}$ has support in $[-\nu,E+\nu]$ and is bounded by 1.
As $dN_{\omega}^N$ and $dN_s$ are positive measures, we have
\begin{equation}
  \label{eq:23}
  \esp(N^N_{s,\omega}(E))
  \leq\esp((\varphi_{E,\nu},dN^N_{\omega}))
  \leq\esp(N^N_{s,\omega}(E+\nu)).
\end{equation}
Hence, by~\eqref{eq:22} and~\eqref{eq:23}, we obtain
\begin{equation*}
  \begin{split}
    N_s(E)&\leq(\varphi_{E,\nu},dN_s)
    =\esp[(\varphi_{E,\nu},dN^N_{\omega})]+
    \left[(\varphi_{E,\nu},dN_s)-
      \esp((\varphi_{E,\nu},dN^N_{\omega}))\right]
    \\&\leq\esp(N^N_{\omega}(E+\nu)) + e^{-\nu^{-\eta/(2\alpha)}}
  \end{split}
\end{equation*}
and
\begin{equation*}
  \begin{split}
    N_s(E)&\geq(\varphi_{E-\nu,\nu},dN_s)
    =\esp[(\varphi_{E-\nu,\nu},dN^N_{\omega})]+
    \left[(\varphi_{E-\nu,\nu},dN_s)-
      \esp((\varphi_{E-\nu,\nu},dN^N_{\omega}))\right] \\&\geq
    \esp(N^N_{\omega}(E-\nu)) - e^{-\nu^{-\eta/(2\alpha)}}
  \end{split}
\end{equation*}
This completes the proof of Lemma~\ref{estapp}.\qed
\subsection{Some Floquet theory}
\label{sec:some-floquet-theory}
To analyze the spectrum of $H^N_{\omega}$, we use some Floquet theory
that we develop now. We denote by $\F:\ L^2([-\pi,\pi]^d)\to
\ell^2(\Z^d)$ the standard Fourier series transform. Then, we have,
for $u\in L^2([-\pi,\pi]^d)$,
\begin{gather*}
  ({\hat H_{\omega}}u)(\theta)=(\F^*H_{\omega}\F u)(\theta)
  =h(\theta)u(\theta)+\sum_{\beta\in\Z^d}\omega_\beta(\Pi_\beta
  u)(\theta) \\ \text{ where }(\Pi_\beta u)(\theta)=
  \frac1{(2\pi)^d}e^{i\beta \theta} \int_{[-\pi,\pi]^d}e^{-i\beta
    \theta}u(\theta)d\theta.
\end{gather*}
Define the unitary equivalence
\begin{equation*}
  \begin{aligned}
    U:\ &L^2([-\pi,\pi]^d)&\to\
    &L^2([-\frac\pi{2N+1},\frac\pi{2N+1}]^d)
    \otimes\ell^2(\Z^d_{2N+1})\\ &\hskip1.8cm u &\mapsto\
    &(Uu)(\theta)= (u_\gamma(\theta))_{\gamma\in\Z^d_{2N+1}}
  \end{aligned}
\end{equation*}
where the $(u_\gamma(\theta))_{\gamma\in\Z^d_{2N+1}}$ are defined by
\begin{equation}
  \label{decfloh}
    u(\theta)=\sum_{\gamma\in\Z^d_{2N+1}}e^{i\gamma\theta}u_\gamma(\theta)\
    \text{ where the functions }(\theta\mapsto
    u_\gamma(\theta))_{\gamma\in\Z^d_{2N+1}}\text{ are
      }\frac{2\pi}{2N+1}\Z^d\text{-periodic.}
\end{equation}
The functions $(u_\gamma)_{\gamma\in\Z^d_{2N+1}}$ are computed easily;
if the Fourier coefficients of $u$ are denoted by $(\hat
u_\gamma)_{\gamma\in\Z^d}$, then, one gets
\begin{equation}
  \label{eq:21}
  u_\gamma(\theta)=\sum_{\beta\in\Z^d}\hat
  u_{\gamma+(2N+1)\beta}e^{i(2N+1)l\beta\theta}.
\end{equation}
The operator $U\F^* H_{\omega}^N \F U^*$ acts on
$L^2([-\frac\pi{2N+1},\frac\pi{2N+1}]^d)\otimes\ell^2(\Z^d_{2N+1})$;
it is the multiplication by the matrix
\begin{equation}
  \label{eq:5}
  M_{\omega}^N(\theta)=H^N(\theta)+ V_\omega^N
\end{equation}
where
\begin{equation}
  \label{eq:32}
  H^N(\theta)=((h_{\beta -\beta'}(\theta)))_{(\beta ,\beta')\in
    (\Z^d_{2N+1})^2}\text{ and }V_\omega^N
  =((\omega_{\beta_1}
  \delta_{\beta_1\beta_1'}\delta_{\beta_20}\delta_{\beta_2'0})
  )_{\substack{(\beta_1,\beta_1')\in
      (\Z^{d_1}_{2N+1})^2\\(\beta_2,\beta_2')\in
      (\Z^{d_2}_{2N+1})^2}}.
\end{equation}
Here, the functions $(h_\gamma)_{\gamma\in\Z^d_{2N+1}}$ are the
components of $h$ decomposed according to~\eqref{decfloh}. The
$(2N+1)^d\times (2N+1)^d$-matrices $H^N(\theta)$ and $V_\omega^N$ are
non-negative matrices.
\par This immediately tells us that the Floquet eigenvalues and
eigenvectors of $H_{\omega}^N$ with Floquet quasi-momentum $\theta$
(i.e. the vectors, $u=(u_\beta )_{\beta \in\Z^d})$, solution to the
problem
\begin{equation*}
  \begin{cases}
    H_{\omega}^Nu&=\lambda u,\\
    u_{\beta +\gamma}&=e^{-i\gamma\theta}u_\beta \text{ for }\beta
    \in\Z^d,\ \gamma\in(2N+1)\Z^d)
  \end{cases}
\end{equation*}
are the eigenvalues and eigenvectors (once extended
quasi-periodically) of the $(2N+1)^d\times (2N+1)^d$ matrix
$M_{\omega}^N(\theta)$. For $E\in\R$, one has
\begin{equation*}
  {\mathcal  N}_{\omega}^N(E)=\int_0^Edn^N_{\omega}(E)
  =\int_{[-\frac{\pi}{2N+1},
    \frac{\pi}{2N+1}]^d}\sharp\{\text{eigenvalues of
    }M_{\omega}^N(\theta)\text{ in }[0,E]\}d\theta.
\end{equation*}
Considering $H$ as $(2N+1)\Z^d$-periodic on $\Z^d$, we see that the
Floquet eigenvalues of $H$ (for the quasi-momentum $\theta$) are
$(h(\theta+\frac{2\pi \gamma}{2N+1}))_{\gamma\in\Z^d_{2N+1}}$; the
Floquet eigenvalue $h(\theta+\frac{2\pi \gamma}{2N+1})$ is associated
to the Floquet eigenvector $u_\gamma(\theta)$, $\gamma\in\Z^d_{2N+1}$
defined by
\begin{equation*}
  u_\gamma(\theta)=\frac1{(2N+1)^{d/2}}(e^{-i(\theta+\frac{2\pi
  \gamma}{2N+1})\beta })_{\beta \in\Z^d_{2N+1}}.
\end{equation*}
In the sequel, the vectors in $\ell^2(\Z^d_{2N+1})$ are given by their
components in the orthonormal basis
$(u_\gamma(\theta))_{\gamma\in\Z^d_{2N+1}}$.  The vectors of the
canonical basis denoted by $(v_l(\theta))_{l\in\Z^d_{2N+1}}$ have the
following components in this basis
\begin{equation*}
  v_l(\theta)=\frac1{(2N+1)^{d/2}}(e^{i(\theta+\frac{2\pi
  \gamma}{2N+1})l})_{\gamma\in\Z^d_{2N+1}}.
\end{equation*}
We define the vectors $(v_l)_{l\in\Z^d_{2N+1}}$ by
\begin{equation*}
  v_l=e^{-il\theta}v_l(\theta)=\frac1{(2N+1)^{d/2}}(e^{i\frac{2\pi
    \gamma l}{2N+1}})_{\gamma\in\Z^d_{2N+1}}.
\end{equation*}
\section{The proof of Theorem~\ref{thr:1}}
\label{sec:stable-edges}
To prove Theorem~\ref{thr:1}, we will use Lemma~\ref{estapp} and the
Floquet theory developed in~\ref{sec:some-floquet-theory}. We will
start with
\subsection{The Floquet theory for constant surface potential}
\label{sec:floq-theory-const}
We consider the operator $H_t^N=H_\omega^N$ where
$\omega=(t)_{\gamma_1\in\Z^{d_1} _{2N+1}}$ is the constant vector and
$t\not=0$. The matrix $M^N_t(\theta)$ defined by~\eqref{eq:5} for
$H_t^N$ takes the form~\eqref{eq:5} where
\begin{equation}
  \label{eq:81}
  V_t^N=t((\delta_{\beta_1\beta_1'}\delta_{\beta_20}
  \delta_{\beta_2'0}))_{\substack{(\beta_1,\beta_1')\in
      (\Z^{d_1}_{2N+1})^2\\(\beta_2,\beta_2')\in
      (\Z^{d_2}_{2N+1})^2}}.
\end{equation}
Our goal is to describe the eigenvalues and eigenfunctions of
$M^N_t(\theta)$. As usual, we write $\theta=(\theta_1,\theta_2)$. By
definition, the operator $H_t^N$ is $\Z^{d_1}\times(2N+1)
\Z^{d_2}$-periodic. It can be seen as acting on $\ell^2(\Z^{d_1},
\ell^2(\Z^{d_2}))$; as such, we can perform a Floquet analysis in the
$\theta_1$-variable as in section~\ref{sec:some-floquet-theory} (in
this case, just a discrete Fourier transform in $\theta_1$) to obtain
that $H_t^N$ is unitarily equivalent to the direct sum over $\theta_1$
in $\T^{d_1}$ of the $2N+1$-periodic operator $H_t^N(\theta_1)$ acting
on $\ell^2(\Z^{d_2})$ defined by the matrix
\begin{equation*}
  H_t^N(\theta_1)=((h(\theta_1;\beta_2-\beta'_2)+
  t\sum_{\gamma_2\in(2N+1)\Z^{d_2}}\delta_{\beta_2\gamma_2}
  \delta_{\beta_2'\gamma_2}))_{(\beta_2,\beta_2')\in(\Z^{d_2})^2}.
\end{equation*}
Here $h(\theta_1;\beta_2)$ is the partial Fourier transform of
$h(\theta_1,\theta_2)$ in the $\theta_2$-variable.\\
For each $\theta_1$, we now perform a Floquet reduction for
$H_t^N(\theta_1)$ to obtain that $H_t^N(\theta_1)$ is unitarily
equivalent to the multiplication by the matrix
\begin{equation*}
  \tilde M^N_t(\theta_1,\theta_2)=
  ((h(\theta_1,\theta_2;\beta_2-\beta'_2)+t\delta_{\beta_20}
  \delta_{\beta_2'0}))_{(\beta_2,\beta_2')\in(\Z^{d_2}_{2N+1})^2}
\end{equation*}
The matrix-valued function $(\theta_1,\theta_2)\mapsto\tilde
M^N_t(\theta_1,\theta_2)$ is $2\pi\Z^{d_1}$-periodic in $\theta_1$ and
$\frac{2\pi}{2N+1}\Z^{d_2}$-periodic in $\theta_2$. It is a rank one
perturbation of the matrix $\tilde M^N_0(\theta_1,\theta_2)$; the
eigenvalues of this matrix are the values $\D
h\left(\theta_1,\theta_2+\frac{2\pi\gamma_2}{2N+1}\right)$.  Let us
for a while order these values increasingly and call them
$(E^N_n(\theta_1,\theta_2,t))_{1\leq n\leq n_N})$ where
$n_N\leq(2N+1)^{d_2}$ (we do not repeat the eigenvalues according to
multiplicity). Assume $t>0$. The standard theory of rank one
perturbations~\cite{MR97c:47008} yields
\begin{Le}
  \label{le:8}
  For $1\leq n\leq n_N$, if $E^N_n(\theta_1,\theta_2,0)$ is an
  eigenvalue of multiplicity $k$, then
  \begin{itemize}
  \item either it is an eigenvalue of multiplicity $k$ for $\tilde
    M^N_t(\theta_1)$;
  \item or it is an eigenvalue of multiplicity $k-1$ for $\tilde
    M^N_t(\theta_1)$ and the interval
    $(E^N_n(\theta_1,0),E^N_{n+1}(\theta_1,\theta_2,0))$ contains
    exactly one simple eigenvalue; this eigenvalue is given by the
    condition
    \begin{equation*}
      t\langle\delta_0,(E-\tilde M^N_0(\theta_1))^{-1}\delta_0\rangle=1;
    \end{equation*}
  \end{itemize}
  Here, we took the convention $E^N_{n_N+1}(\theta_1,\theta_2,0)
  =+\infty$. One has a symmetric statement for $t<0$.
\end{Le}
\noindent For $1\leq n\leq n_N$, let $(\varphi_{n,j}^N(\theta_1,\theta_2,t))
_{1\leq j\leq j_n}$ denote orthonormalized eigenvectors associated to
the eigenvalue $E^N_n(\theta_1,\theta_2,t)$ where $j_n$ denotes its
multiplicity.\\
In the sequel, it will be convenient to reindex the eigenvalues and
eigenfunctions of the matrix $\tilde M^N_t(\theta_1,\theta_2)$ as
$(E^N_{\gamma_2}( \theta_1,\theta_2,t))_{\gamma_2\in\Z^{d_2}_{2N+1}}$
and $(\varphi_{\gamma_2}^N(\theta_1,\theta_2,t))_{\gamma_2\in\Z^{d_2}
  _{2N+1}}$. Clearly, the functions $(\theta_1,\theta_2)\mapsto
E^N_{\gamma_2} (\theta_1,\theta_2,t)$ and $(\theta_1,\theta_2)\mapsto
\varphi_{\gamma_2}^N (\theta_1,\theta_2,t)$ can be chosen to be
$2\pi\Z^{d_1}$-periodic in $\theta_1$ and
$\frac{2\pi}{2N+1}\Z^{d_2}$-periodic in $\theta_2$.\\
Let us now show the
\begin{Le}
  \label{le:4}
  The eigenvalues of $M^N_t(\theta)$ are the values
  $\{E_{\gamma_1,\gamma_2}(\theta_1,\theta_2,t);\ \gamma_1\in
  \Z^{d_1}_{2N+1},\ \gamma_2\in\Z^{d_2}_{2N+1}\}$ where
  \begin{equation}
    \label{eq:1}
    E_{\gamma_1,\gamma_2}(\theta_1,\theta_2,t)=
    E^N_{\gamma_2}\left(\theta_1+\frac{2\pi\gamma_1}{2N+1},
      \theta_2,t\right)
  \end{equation}
  A normalized eigenfunction associated to the eigenvalue
  $E^N_{\gamma_2}\left(\theta_1+\frac{2\pi\gamma_1}{2N+1},\theta_2,
    t\right)$ is the vector
  \begin{equation}
    \label{eq:77}
    v_{\gamma_1,\gamma_2}(\theta_1,\theta_2,t)
    :=(2N+1)^{-d_1/2}\left(e^{-i\beta_1(\theta_1
        +\frac{2\pi\gamma_1}{2N+1})} \varphi_{\gamma_2}^N
      \left(\theta_1+\frac{2\pi\gamma_1}{2N+1},\theta_2,t\right)
    \right)_{\beta_1\in\Z^{d_1}_{2N+1}},
  \end{equation}
  i.e. the vector of components
  \begin{equation}
    \label{eq:78}
    (2N+1)^{-d_1/2}\left(e^{-i\beta_1(\theta_1+
        \frac{2\pi\gamma_1}{2N+1})}
      c_{\gamma_2}^{\beta_2}\left(\theta_1+
        \frac{2\pi\gamma_1}{2N+1}\right)
    \right)_{\substack{\beta_1\in\Z^{d_1}_{2N+1}\\
        \beta_2\in\Z^{d_2}_{2N+1}}}
  \end{equation}
  if $\varphi_{\gamma_2}^N(\theta_1,t)$ has components
  $(c_{\gamma_2}^{\beta_2}(\theta_1))_{\beta_2\in\Z^{d_2}_{2N+1}}$.\\
  The vectors $(v_{\gamma_1,\gamma_2}(\theta_1,\theta_2,t))_
  {\substack{\gamma_1\in\Z^{d_1}_{2N+1}\\
      \gamma_2\in\Z^{d_2}_{2N+1}}}$ form an orthonormal basis of
  $\ell^2(\Z^{d_1}_{2N+1}\times\Z^{d_2}_{2N+1})$.
\end{Le}
\noindent{\bf Proof.} Orthonormality is easily checked using the fact
that the vectors $(\varphi_{\gamma_2}^N(
\theta_1,\theta_2,t))_{\gamma_2\in\Z^{d_2}_{2N+1}}$ form an
orthonormal basis.\\
Let us now check that $v_{\gamma_1,\gamma_2}(\theta_1,\theta_2,t)$
satisfies the eigenvalue equation for $M^N_t(\theta)$ and $\D
E_{\gamma_1,\gamma_2}(\theta_1,\theta_2,t)$ given in~\eqref{eq:1}.
Therefore, first note that the matrix $M^N_t(\theta)$ is nothing but
the multiplication operator by the matrix-valued function $\tilde
M^N_t(\theta_1)$ to which one has applied the Floquet reduction of in
the $\theta_1$-variable. Hence, by~\eqref{eq:21}, the matrix elements
of $M^N_t(\theta)$ given by~\eqref{eq:32} satisfy, for
$\beta_1\in\Z^{d_1}_{2N+1}$,
\begin{equation}
  \label{eq:67}
  \tilde M^N_t\left(\theta_1+\frac{2\pi\gamma_1}{2N+1}\right)
  e^{-i\beta_1(\theta_1+\frac{2\pi\gamma_1}{2N+1})}=
  \sum_{\beta'_1\in\Z^{d_1}_{2N+1}}m_{\beta_1-\beta'_1}(\theta_1)
  e^{-i\beta'_1(\theta_1+\frac{2\pi\gamma_1}{2N+1})}
\end{equation}
Both sides in this equality are matrices acting on
$\ell^2(\Z^{d_2}_{2N+1})$, the matrices $m_{\beta_1-\beta'_1}(\theta)$
being defined as
\begin{equation*}
  m_{\beta_1-\beta'_1}(\theta)=
  ((h_{\beta_1-\beta'_1,\beta_2-\beta'_2}(\theta)))
  _{(\beta_2,\beta'_2)\in(\Z^{d_2}_{2N+1})^2}.
\end{equation*}
If we now apply both sides of equation~\eqref{eq:67} to the vector
$\varphi_{\gamma_2}^N\left(\theta_1+\frac{2\pi\gamma_1}{2N+1},
  t\right)$, we obtain, for $\beta_1\in\Z^{d_1}_{2N+1}$,
\begin{equation*}
  \begin{split}
    &\sum_{\beta'_1\in\Z^{d_1}_{2N+1}}h_{\beta_1-\beta'_1}(\theta_1)
    e^{-i\beta'_1(\theta_1+\frac{2\pi\gamma_1}{2N+1})}
    \varphi_{\gamma_2}^N\left(\theta_1+\frac{2\pi\gamma_1}{2N+1},
      \theta_2,t\right)\\ &\hskip1cm=\tilde
    M^N_t\left(\theta_1+\frac{2\pi\gamma_1}{2N+1}\right)
    e^{-i\beta_1(\theta_1+\frac{2\pi\gamma_1}{2N+1})}
    \varphi_{\gamma_2}^N\left(\theta_1+
      \frac{2\pi\gamma_1}{2N+1},t\right)\\
    &\hskip1cm=E^N_{\gamma_2}\left(\theta_1+\frac{2\pi\gamma_1}
      {2N+1},\theta_2,t\right)
    e^{-i\beta_1(\theta_1+\frac{2\pi\gamma_1}{2N+1})}
    \varphi_{\gamma_2}^N\left(\theta_1+
      \frac{2\pi\gamma_1}{2N+1},\theta_2,t\right)
  \end{split}
\end{equation*}
This is nothing but to write
\begin{equation*}
  M^N_t(\theta)v_{\gamma_1,\gamma_2}(\theta_1,\theta_2,t)=
  E^N_{\gamma_2}\left(\theta_1+\frac{2\pi\gamma_1}{2N+1},\theta_2,t\right)
  v_{\gamma_1,\gamma_2}(\theta_1,\theta_2,t).
\end{equation*}
This completes the proof of Lemma~\ref{le:4}.\qed\\
In the course of the proof of Theorem~\ref{thr:1}, we will use the
\begin{Le}
  \label{le:5}
  Fix $t$ such that $t>0$ if $d_2=1,2$ and $1+tI_{\infty}>0$ if
  $d_2\geq3$. Then, for $\rho>2$, there exists $C>0$ such that, for
  $N\geq E^{-\rho}$ and $E$ sufficiently small, the eigenvalues of
  $M^N_t$ satisfy
  \begin{equation}
    \label{eq:68}
    E_{\gamma_1,\gamma_2}(\theta_1,\theta_2,t)\leq E\Longrightarrow
    \left(\frac{1+|\gamma_1|}{2N+1}\right)^2 \leq C E
  \end{equation}
\end{Le}
\noindent{\bf Proof.} When $t$ is positive,~\eqref{eq:68} is clear
by Lemmas~\ref{le:4} and~\ref{le:8}, that is, by the intertwining of
the eigenvalues of $M^N_0(\theta)$ and $M^N_t(\theta)$, and as the
eigenvalues of $M^N_0(\theta)$ are the values $\D h\left(
  \theta_1+\frac{2\pi\gamma_1}{2N+1},\theta_2+\frac{2\pi\gamma_2}{2N+1}
\right)$ which satisfy~\eqref{eq:68} as $h(\theta)\geq C|\theta|^2$.
\par Assume now that $d_2\geq3$ and $t$ satisfies
$1+tI_{\infty}>0$. To complete the proof of Lemma~\ref{le:5}, by
Lemma~\ref{le:4}, it is then enough to prove that, there exists $C>0$
such that for, one has
\begin{equation*}
    |\theta_1|^2>CE\Longrightarrow \forall\gamma_2,\
     E^N_{\gamma_2}(\theta_1,\theta_2,t)>E.
\end{equation*}
By the intertwining properties and the properties of $h$, this is
clear except for the lowest of the $(E^N_{\gamma_2})_{\gamma_2}$.
Assume now that $|\theta_1|^2 \geq E$. Then, by our assumptions on the
behavior of $h$ near its minimum, for some $C>0$, one has that
$(\theta_1,e)\mapsto I(\theta_1,e)$ is real analytic in
$\{|\theta_1|^2\geq E\}\times\{|e|\leq E/C\}$. Hence, using a standard
estimate for Riemann sums, we get that, for $|\theta_1|^2\geq E$ and
$|e|\leq E/C$,
\begin{equation*}
  1+t\langle\delta_0,(\tilde M^N_0(\theta_1)-e)^{-1}\delta_0\rangle
  = 1+tI(\theta_1,e)+O(E^{-2}E^{\rho})
\end{equation*}
So that, as $1+tI_\infty>0$, for $E$ sufficiently small, the equation
$1+t\langle\delta_0,(\tilde M^N_0(\theta_1)-e)^{-1}\delta_0\rangle=0$
has no solution for $|\theta_1|^2\geq E$ and $|e|\leq E/C$. By the
above discussion, this implies that, all the
$E^N_{\gamma_2}(\theta_1,\theta_2,t)$ lie above $E/C$. This completes
the proof of Lemma~\ref{le:5}.\qed
\subsection{The proof of Theorem~\ref{thr:1}}
\label{sec:proof-theor-refthr:1}
We now have all the tools necessary to prove Theorem~\ref{thr:1}.
Notice that, as $\overline{\omega}>\omega_-$, as
$1+\omega_-I_{\infty}\geq0$, we know that
$1+\overline{\omega}I_{\infty}>0$. So that the asymptotics for
$N_s^{\overline{\omega}}(E)$ are given by
\begin{equation*}
  N_s^{\overline{\omega}}(E)\equ_{E\to0^+} C(\overline{\omega})\cdot
  f(E).
\end{equation*}
The precise value of the constant $C(\overline{\omega})$ and of the
function $f(E)$ are given in Theorem~\ref{thr:3}. The constant
$C(\overline{\omega})$ is a continuous function of
$\overline{\omega}$; and, for any $c\in\R$, the function $f(E)$
satisfies $f(E+cE^2)\sim f(E)$ when $E\to0$; moreover, $f$ is at most
polynomially small in $E$.  All these facts will be useful.
\vskip.2cm\noindent We start with the proof of~\eqref{eq:75}. We will
use Lemma~\ref{estapp}. As above, fix $N$ large but not too large, say
$N\sim E^{-\rho}$ for some large $\rho$.  Fix $\delta>0$ small.
Consider the matrix $M^N_{\overline{\omega}+\delta}(\theta)$ obtained
by the Floquet reduction of $H^N+(\overline{\omega}+\delta)\Pi_0^2$.
Let $\mathcal{H}_\delta^N(E,\theta)$ be the spectral space of
$M^N_{\overline{\omega}+\delta}(\theta)$ associated the eigenvalues
less that $E$. Then, we prove that
\begin{Le}
  \label{le:6}
  Fix $\delta>0$, $\rho>2$ and $\alpha\in(0,1/2)$. For $N\sim
  E^{-\rho}$ and $E$ sufficiently small, with a probability at least
  $1-e^{E^{-\alpha}}$, for all $\theta$ and all
  $\varphi\in\mathcal{H}_\delta^N(E,\theta)$, one has
  \begin{equation*}
    \langle M^N_\omega(\theta)\varphi,\varphi\rangle\leq
    E\Vert\varphi\Vert^2.
  \end{equation*}
\end{Le}
\noindent This lemma immediately implies the desired lower bound.
Indeed, it implies that, for $N\sim E^{-\rho}$, with a probability at
least $1-e^{E^{-\alpha}}$, one has
\begin{equation*}
  \begin{split}
    N_{\overline{\omega}+\delta}^N(E)
    &=\int_{[-\frac{\pi}{2N+1},\frac{\pi}{2N+1}]^d}\sharp
    \{\text{eigenvalues of }M_{\overline{\omega}+\delta}^N(\theta)
    \text{ in }[0,E]\}d\theta\\&\leq\int_{[-\frac{\pi}{2N+1},
      \frac{\pi}{2N+1}]^d}\sharp\{\text{eigenvalues of
    }M_{\omega}^N(\theta)\text{ in }[0,E]\}d\theta\\
    &=N_{\omega}^N(E) \end{split}
\end{equation*}
Taking the expectation of both side, and using~\eqref{estpre} (and the
fact that the number of eigenvalues of $M^n_\omega(\theta)$ and
$M^N_{\overline{\omega}+\delta}(\theta)$ are bounded by $(2N+1)^d$),
we obtain
\begin{equation*}
  N_s^{\overline{\omega}+\delta}(E-E^2)-C E^{d\rho}e^{-E^{-\alpha}}\leq
  N_s(E)
\end{equation*}
Considering the remarks made above, we obtain
\begin{equation*}
  C(\overline{\omega})\leq\liminf_{E\to0^+}\frac{N_s(E)}{f(E)}.
\end{equation*}
As $C(\overline{\omega})$ has the same sign as $\overline{\omega}$,
this completes the proof of~\eqref{eq:75}.
\vskip.2cm\noindent{\bf Proof of Lemma~\ref{le:6}.} Pick $E$ small and
$\varphi\in\mathcal{H}_\delta^N(E,\theta)$. Then, by Lemma~\ref{le:5},
$\varphi$ can be expanded as
\begin{equation*}
  \varphi=\sum_{\substack{|\gamma_1|\leq CE^{1/2}N\\
    \gamma_2\in\Z^{d_2}_{2N+1}}} a_{\gamma_1,\gamma_2}
  v_{\gamma_1,\gamma_2}(\theta,t)
\end{equation*}
where the vectors $(v_{\gamma}(\theta))_\gamma$ are given
by~\eqref{eq:77} and~\eqref{eq:78}. Using these equations, we compute
\begin{equation}
  \label{eq:70}
    \langle V_\omega^N\varphi,\varphi\rangle=
    \sum_{\beta_1\in\Z^{d_1}_{2N+1}}\omega_{\beta_1}|A_{\beta_1}|^2
\end{equation}
where
\begin{equation}
  \label{eq:88}
  A_{\beta_1}=\frac{1}{(2N+1)^{d_1/2}}
  \sum_{|\gamma_1|\leq CE^{1/2}N}
  e^{i\frac{2\pi\beta_1\cdot\gamma_1}{2N+1}}c_{\gamma_1}
  \text{ and }c_{\gamma_1}=\sum_{\gamma_2\in\Z^{d_2}_{2N+1}}
  a_{\gamma_1,\gamma_2}\left\langle\delta_0,\varphi_{\gamma_2}^N
    \left(\theta_1+\frac{2\pi\gamma_1}{2N+1},\theta_2,t\right)
  \right\rangle
\end{equation}
So the vector $(A_{\beta_1})_{\beta_1}$ is the discrete Fourier
transform of the vector $c=(c_{\gamma_1})_{\gamma_1}$ supported in a
ball of radius $C E^{1/2}N$. To estimate this Fourier transform, we
used the following result
\begin{Le}[\cite{Kl:01a}]
  \label{lemcle}
  Assume $N$, $L$, $K$, $K'$ $L'$ are positive integers such that
  \begin{itemize}
  \item $2N+1=(2K+1)(2L+1)=(2K'+1)(2L'+1)$
  \item $K<K'$ and $L'<L$.
  \end{itemize}
  Pick $a=(a_n)_{n\in\Z^d_{2N+1}}\in\ell^2(\Z^d_{2N+1})$ such that,
  \begin{equation*}
    \text{for }\vert n\vert>K,\ a_n=0.
  \end{equation*}
  Then, there exists $\tilde a\in\ell^2(\Z^d_{2N+1})$ such that
  \begin{enumerate}
  \item \label{01} $\D\Vert a-\tilde a\Vert_{\ell^2(\Z^d_{2N+1})}\leq
    C_{K,K'}\Vert a\Vert_{\ell^2(\Z^d_{2N+1})}$ where $\D
    C_{K,K'}\asymp_{K/K'\to0}K/K'$.
  \item \label{02} write $\tilde a=(\tilde a_j)_{j\in\Z^d_{2L+1}}$;
    for $l'\in\Z^d_{2L'+1}$ and $k'\in\Z^d_{2K'+1}$, we have
    \begin{equation*}
      \sum_{j\in\Z^d_{2L+1}}\tilde a_j
      e^{i\frac{2\pi j\cdot(l'+k'(2L'+1))}{2N+1}}=
      \sum_{j\in\Z^d_{2L+1}}\tilde a_j
      e^{i\frac{2\pi j\cdot k'}{2K'+1}}.
    \end{equation*}
  \item \label{03} $\D\Vert a\Vert_{\ell^2(\Z^d_{2N+1})}= \Vert\tilde
    a\Vert_{\ell^2(\Z^d_{2N+1})}$.
  \end{enumerate}
\end{Le}
\noindent This lemma is a quantitative version of the Uncertainty
Principle; it says that, if a vector is localized in a small
neighborhood of $0$ (here, of size $K/N$), up to a small error
$\delta$, its Fourier transform is constant over cube of size
$N/(\delta K)$.
\par To apply Lemma~\ref{lemcle}, we pick $N$ such that
$(2N+1)=(2K+1)(2L'+1)(2M+1)$ where $K\geq CE^{1/2}N$; this is possible
as $N\sim E^{-\rho}$ with $\rho$ large; we pick for example, $L'\sim
CE^{-(1-\nu)/2}$ and $L'\sim CE^{-\nu/2}$ (for some fixed
$0<\nu<1$).\\
We apply Lemma~\ref{lemcle} to the vector
$c=(c_{\gamma_1})_{\gamma_1}$ defined in~\eqref{eq:88}; by
Lemma~\ref{lemcle}, there exists $\tilde c=(\tilde
c_{\gamma_1})_{\gamma_1}$ so that, if we set
\begin{equation*}
  \tilde A_{\beta_1}=\frac{1}{(2N+1)^{d_1/2}}
  \sum_{\gamma_1\in\Z^{d_1}_{2N+1}}
  e^{i\frac{2\pi\beta_1\cdot\gamma_1}{2N+1}}c_{\gamma_1}
\end{equation*}
then, for $\gamma'_1\in\Z^d_{2L'+1}$ and $\beta'_1\in\Z^d_{2K'+1}$, we
have
\begin{equation}
  \label{eq:80}
  \tilde A_{\gamma'_1+\beta'_1(2L'+1)}=\tilde A_{\beta'_1(2L'+1)}.
\end{equation}
Fix $\eta>0$ small to be chosen later. We replace $A$ by $\tilde A$
in~\eqref{eq:70} and use the boundedness of the random variables to
obtain
\begin{equation*}
    \langle V_\omega^N\varphi,\varphi\rangle\leq
    (1+\eta)\sum_{\beta_1\in\Z^{d_1}_{2N+1}}\omega_{\beta_1}|\tilde
    A_{\beta_1}|^2+\frac{C}{\eta}\Vert A-\tilde A\Vert^2
\end{equation*}
Using~\eqref{eq:80} and points (1) and (3) of Lemma~\ref{lemcle}, we
get that
\begin{equation*}
    \langle V_\omega^N\varphi,\varphi\rangle\leq
    \sum_{\beta'_1\in\Z^{d_1}_{2K'+1}}
    \left[\frac{C}{\eta}E^{\nu/2}+\frac{1}{(2L'+1)^{d_1}}
    \left(\sum_{\gamma'_1\in\Z^{d_1}_{2L'+1}}(1+\eta)
      \omega_{\gamma'_1+\beta'_1(2L'+1)}\right)\right]
    (2L'+1)^{d_1}|\tilde A_{\beta'_1}(2L'+1)|^2
\end{equation*}
Pick $\eta$ such that $\eta\cdot\omega_+<\delta/4$ and $E$
sufficiently small that $\frac{C}{\eta}E^{\nu/2}<\delta/4$. We then
obtain
\begin{equation}
    \label{eq:79}
    \langle V_\omega^N\varphi,\varphi\rangle\leq
    \sum_{\beta'_1\in\Z^{d_1}_{2K'+1}}
    \left[\delta/2+\frac{1}{(2L'+1)^{d_1}}
    \left(\sum_{\gamma'_1\in\Z^{d_1}_{2L'+1}}
      \omega_{\gamma'_1+\beta'_1(2L'+1)}\right)\right]
    (2L'+1)^{d_1}|\tilde A_{\beta'_1}(2L'+1)|^2
\end{equation}
Now, if $\omega$ satisfies
\begin{equation*}
  \forall\beta'_1\in\Z^{d_1}_{2K'+1},\
    \frac{1}{(2L'+1)^{d_1}}\sum_{\gamma'_1\in\Z^{d_1}_{2L'+1}}
      \omega_{\gamma'_1+\beta'_1(2L'+1)}
    \leq\overline{\omega}+\delta/2
\end{equation*}
then,~\eqref{eq:79} gives
\begin{equation*}
  \langle V_\omega^N\varphi,\varphi\rangle\leq
    (\overline{\omega}+\delta)
    \sum_{\beta'_1\in\Z^{d_1}_{2K'+1}}
    (2L'+1)^{d_1}|\tilde A_{\beta'_1}(2L'+1)|^2=
    \langle V^N_{\overline{\omega}+\delta}\varphi,\varphi\rangle
\end{equation*}
where $V^N_t$ is defined in~\eqref{eq:81}.  Here, we have used the
points (2) and (3) of Lemma~\ref{lemcle}, and the
definition~\eqref{eq:88} of the
vector $c=(c_{\gamma_1})_{\gamma_1}$.\\
Summing all this up, we have proved
\begin{Le}
  \label{le:7}
  Pick $0<\nu<1$. Pick $N$ as described above. For $E$ sufficiently
  small, the probability that, for all $\theta$ and all
  $\varphi\in\mathcal{H}_\delta^N(E,\theta)$, one has
  \begin{equation*}
    \langle M^N_\omega(\theta)\varphi,\varphi\rangle\leq
    E\Vert\varphi\Vert^2.
  \end{equation*}
  is larger than the probability of the set
  \begin{equation*}
    \left\{\omega;\ \forall\beta'_1\in\Z^{d_1}_{2K'+1},\
      \frac{1}{(2L'+1)^{d_1}}\sum_{\gamma'_1\in\Z^{d_1}_{2L'+1}}
      \omega_{\gamma'_1+\beta'_1(2L'+1)}
    \leq\overline{\omega}+\delta/2\right\}
  \end{equation*}
\end{Le}
\noindent The probability of this event is estimated by the usual
large deviation estimates (see e.g.~\cite{MR98m:60001,De-Ze:92}). This
completes the proof of Lemma~\ref{le:6}.\qed
\section{The fluctuating edges}
\label{sec:fluctuating-edges}
In this section, we investigate the behavior of the density of surface
states $N_s(E)$ at the bottom $E_0$ of the spectrum of $H_{\omega}$ in
the case when $E_0<\inf\sigma(H)=0$. As we saw in
Section~\ref{sec:stable-edge}, this is always the case for dimension
$d_2=1$ or $d_2=2$ and it holds in arbitrary dimensions if the support
of common distribution $P_0$ of the $\omega_{\gamma_1}$ has a
sufficiently negative part. Thus, we are looking at a fluctuation edge
as described in Section~\ref{sec:fluctuation-edge}. Due to the
symmetry of the problem we may, of course, consider the top of the
spectrum in an analogous way.
\subsection{A reduced Hamiltonian}
\label{sec:reduced-hamiltonian}
In the present situation it is convenient to think of the Hilbert
space $\ell^2(\Z^{d_1+d_2})$ as a direct some of
$\ell^2(\Z^{d_1}\times\{0\})=:\mathcal{H}_b$ and
$\ell^2(\Z^{d_1+d_2}\setminus\Z^{d_1}\times\{0\})=:\mathcal{H}_s$, the
indices referring to ``bulk'' and ``surface'' respectively
(see~\cite{MR2001m:47143} whose notations we follow).  According to
the decomposition $\mathcal{H}=\mathcal{H}_S\oplus \mathcal{H}_b$ we
can write any operator $A$ on $\mathcal{H}$ as a matrix
\begin{equation*}
    A = \left[\begin{array}{cc}
        A_{ss} & A_{sb} \\ A_{bs} & A_{bb}
        \end{array}\right]
\end{equation*}
where $A_{ss}$ and $A_{bb}$ act on $\mathcal{H}_s$ and $\mathcal{H}_b$
respectively and $A_{sb}: \mathcal{H}_b\rightarrow\mathcal{H}s$,
$A_{bs}: \mathcal{H}_s\rightarrow\mathcal{H}_b$ ``connect'' the two
Hilbert spaces $\mathcal{H}_s$ and $\mathcal{H}_b$.  The bounded
operator $A$ is symmetric if $A^*_{ss}=A_{ss}$, $A^*_{bb}=A_{bb}$ and
$A^*_{sb}=A_{bs}$.  In the case of our random Hamiltonian $H_\omega$
we have: $(H_\omega)_{ss}=(H_0)_{ss}+V_\omega$ while
$(H_\omega)_{bb}=(H_0)_{bb}$ and $H_{sb}$ as well as $H_{bs}$ are
independent of the randomness. Moreover, by assumption,
$(H_\omega)_{bb}\geq 0$, while $\inf\sigma (H_\omega)<0$.\par\noindent
Consequently, the operator $\left( (H_0)_{bb}-E\one_{bb}\right)^{-1}$
exists for all $E<0$ and the operator
\begin{equation*}
    G_s(E):=(H_0)_{ss}+V_\omega -
    H_{sb}\left( (H_0)_{bb}-E\one_{bb}\right)^{-1}H_{bs}-E\one_{ss}
\end{equation*}
the so called resonance function is well defined.  The operator
$G_s(E)$ is a sort of a reduced Hamiltonian. Its inverse plays the
role of a resolvent. It is not hard to show that the set
$\mathcal{R}(H_\omega)=\{E\in ]-\infty ,0[ ; 0\in\sigma (G_s(E))\}$
(the resonant spectrum) agrees with the negative part of
$\sigma(H_\omega)$. See Prop.1.2 in~\cite{MR2001m:47143} for details.
For later reference, we state this as a lemma:
\begin{Le}
  \label{le:w1}
  For $E<0$, $E$ is an eigenvalue of $H_\omega$ if and only if $0$ is
  an eigenvalue at $G_s(E)$. Moreover the multiplicities agree.
\end{Le}
\noindent In fact a little linear algebra proves that, for block
matrices, we have
\begin{equation}
  \label{eq:w1}
  \left(\begin{array}{cc}
      A & B \\ C & D
    \end{array}\right)^{-1}
  =
  \left(\begin{array}{cc}
      (A-BD^{-1}C)^{-1} & -A^{-1}B(D-CA^{-1}B)^{-1} \\
      -D^{-1}C(A+BD^{-1}C)^{-1} & (D-CA^{-1}B)^{-1}
    \end{array}\right)
\end{equation}
when all the terms make sense.\\
We denote by $N(A,E)$ the number of eigenvalues (counted according to
multiplicity) of the operator $A$ below $E$.  For $\Lambda_L=[-L,L]^d$
we set $(H_{\omega, L})_{ij}=(H_\omega)_{ij}$ if $i,j\in \Lambda_L$
and $(H_{\omega, L})_{ij}=0$ otherwise.  For energies $E$ below zero
the integrated density of surface states of $H_\omega$ is given by
\begin{equation*}
    N_s(E)=\lim_{L\to\infty}\frac{1}{(2L+1)^{d_1}}N(H_{\omega ,L},E).
\end{equation*}
Defining
\begin{equation*}
    G^L_s(E)=(H_{\omega
    ,L})_{ss}-(H_L)_{sb}((H_{L})_{bb}-E\one_{bb})^{-1}
    (H_L)_{bs}-E\one_{ss}.
\end{equation*}
We have, as above, that $E<0$ is an eigenvalue of $H_{\omega ,L}$ if
and only if $0$ is an eigenvalue of $G^L_s(E)$.\\
In the following, we will express the density of surface states
$N_s(E)$ (for $E<0$) in terms of the operators $G^L_s(E)$.
\begin{Le}
  \label{le:w2}
  The eigenvalues $\rho_n(E)$ of $G^L_s(E)$ are continuous and
  decreasing functions of $E$ (for $E<0$).
\end{Le}
\noindent{\bf Proof of Lemma~\ref{le:w2}.} Continuity is obvious from
the explicit form of the entries of the (finite-dimensional) matrix
$G^L_S(E)$. In the following calculations we omit the superscript $L$.
Let $0>E_2>E_1$, then
\begin{equation*}
  \begin{split}
    G_s(E_1)-G_s(E_2)&=-
    H_{sb}((H_{bb}-E_1)^{-1}-(H_{bb}-E_2)^{-1})H_{bs} -(E_1-E_2) \\
    &=(E_2-E_1)H_{sb}((H_{bb}-E_1)^{-1}
    (H_{bb}-E_2)^{-1})H_{bs}+(E_2-E_1).
  \end{split}
\end{equation*}
Since $E_1,E_2<0$ the operator $(H_{bb}-E_1)^{-1}(H_{bb}-E_2)^{-1}$ is
positive, the operator $G_s(E_1)-G_s(E_2)$ is positive.\hfill\qed
\vskip.2cm\noindent
\begin{Pro}\label{pro:w1}
  For $E<0$:
  \begin{equation*}
    N(H_{\omega,L},E)=N(G^L_s(E),0).
  \end{equation*}
\end{Pro}
\noindent{\bf Proof of Proposition~\ref{pro:w1}.} For $E$ sufficiently
negative, $G_s(E)$ is a positive operator. Let us now increase $E$
(toward $E=0$). Then, $E$ is an eigenvalue of $H_{\omega, L}$ if one
of the eigenvalues of $G^L_s(E)$ passes through zero and becomes
negative.\hfill\qed
\vskip.2cm\noindent It follows from this proposition that (for $E<0$)
\begin{equation*}
    N_s(E)=\lim_{L\to\infty} N(G^L_s(E),0)
\end{equation*}
$G^L_s(E)$ depends on $E$ in a rather complicated way through the
resonance function.  We will therefore approximate $G^L_s(E)$ by an
operator with much simpler dependence on $E$ in the following way:\\
let $E_0=\inf\sigma(H_\omega)$ then we set:
\begin{equation*}
    \tilde{G}^L_s(E)=(H_{\omega,L})_{ss}-(H_L)_{sb}((H_L)_{bb}-E_0)^{-1}(H_L)_{bs}
    -E
\end{equation*}
This operator should give a good estimate for the eigenvalues of
$H_\omega$ near $E_0$, in fact:
\begin{Le}
  \label{le:w3}
  For $E_0<E<0:$
        \[ N(\tilde{G_s}(E),0)\leq N(G_s(E),0) \]
\end{Le}
\noindent{\bf Proof of Lemma~\ref{le:w3}.}
\begin{equation*}
    \tilde{G_s}(E)-G_s(E)=(E-E_0)H_{sb}\left((H_{bb}-E)^{-1}(H_{bb}-E_0)^{-1}\right)
    H_{bs}.
\end{equation*}
So
\begin{equation*}
  \tilde{G^L_s}(E)\geq G^L_s(E).
\end{equation*}
\qed
\vskip.2cm\noindent For a bound in the other direction we observe
that:
\begin{Le}
  \label{le:w4}
  For $E_0\leq E\leq E_1<0$ we have
  \begin{equation*}
    \tilde{G_s}(E)-G_s(E)\leq C (E-E_0)
  \end{equation*}
\end{Le}
\noindent
\textbf{Remark:} The constant $C$ in the above estimate depends on
$E_0$ and $E_1$.\\
{\bf Proof of Lemma~\ref{le:w4}.}
\begin{equation*}
  \begin{split}
    \tilde{G}_s(E)-G_s(E)
    & = (E-E_0)H_{sb}((H_{bb}-E)^{-1}(H_{bb}-E_0)^{-1})H_{bs} \\
    & \leq (E-E_0)H_{sb}((H_{bb}-E_1)^{-1}(H_{bb}-E_0)^{-1})H_{bs} \\
    & \leq C (E-E_0).
  \end{split}
\end{equation*}
Here, we used that
\begin{equation*}
    (H_{bb}-E)^{-1}\leq (H_{bb}-E_1)^{-1}.
\end{equation*}
\hfill\qed\\
Summarizing, we have got:
\begin{Pro}
  \label{pro:w2}
  There is a constant $C$, such that for $E_0\leq E\leq E_0/2<0$
  \begin{equation*}
    N(\tilde{G}_s(E),0)\leq N(H_\omega,E)\leq N(\tilde{G}_s(E)-C(E-E_0),0).
  \end{equation*}
\end{Pro}
\noindent The advantage of having $\tilde{G}_s(E)$ rather than
$G_s(E)$ lies in the fact that $\tilde{G}_s(E)$ depends linearly on
$E$, in fact:
\begin{equation*}
  \begin{split}
    G_s(E) & =H_{ss}-H_{sb}(H_{bb}-E_0)^{-1}H_{bs}+V_\omega -E \\
    & = \tilde{H}+V_\omega-E
  \end{split}
\end{equation*}
where $\tilde{H}$ is the operator
\begin{equation*}
    \tilde{H}=H_{ss}-H_{sb}(H_{bb}-E_0)^{-1}H_{bs}.
\end{equation*}
This operator is of a similar form as the Hamiltonian $H$, however it
acts on $\ell^2(\Z^{d_1})$, i.e. on the surface only where the random
potential $V_\omega$ lives. The price to pay is the complicated
looking ``bulk term'' $H_{sb}(H_{bb}-E_0)^{-1}H_{bs}$.\\
Nevertheless, $\tilde{H}$ is still a Toeplitz operator and it is not
too hard to compute its symbol, i.e. its Fourier representation.\\
In fact, a look at formula~\eqref{eq:w1} shows that
\begin{equation}
  \label{eq:w2}
    \tilde{H}=\left[\left((H-E_0)^{-1}\right)_{ss}\right]^{-1}+E_0.
\end{equation}
Consequently the symbol of $\tilde{H}$ is given by:
\begin{equation*}
    \tilde{h}(\theta_1)=\left(\int\frac{1}{h(\theta_1
    ,\theta_2)-E_0}\;d\theta_2\right)^{-1} + E_0.
\end{equation*}
We summarize these results in a theorem:
\begin{Th}
  \label{thr:10}
  Let $H_\omega=H+V_\omega$ as in~\eqref{eq:8} satisfying assumption
  (H1).  Assume moreover, that $E_0=\inf\sigma(H_\omega)\leq 0$.
  Define $\tilde{H}_\omega=\tilde{H}+\tilde{V}_\omega$ as
  in~\eqref{eq:w2} and let $N_s({H_\omega},E)$ be the integrated
  density of surface states of $H_\omega$ and $N(\tilde{H}_\omega,E)$
  the integrated density of states for $\tilde{H}_\omega$. Then
    \begin{equation*}
        \lim_{E\searrow E_0}\frac{\ln|\ln N_s(H_\omega,E)|}{\ln (E -
        E_0)}
        = \lim_{E\searrow E_0}\frac{\ln|\ln
        N(\tilde{H}_\omega,E)|}{\ln (E - E_0)}
    \end{equation*}
    where the equality should be interpreted in the following way: if
    one of the sides exists so does the other one and they agree.
\end{Th}
\noindent In other words, the Lifshitz exponent for the density of
\textsl{surface} states of $H_\omega$ and and the Lifshitz exponent
for the density of states for $\tilde{H}_\omega$ agree.
\subsection{Lifshitz tails}
\label{sec:lifshitz-tails}
In this section we investigate the integrated density of surface
states $N_s(E)$ for the operator $H_\omega=H+V_\omega$ acting on
$\ell^2(\Z^{d_1}\times\Z^{d_2})$. We assume throughout that $E_0=\inf
\sigma(H_\omega)$ is (strictly) negative and $E<0$.
\par By the previous section the investigation of $N_s(E)$ for $E$
near $E_0$ can be reduced to estimates for the integrated density of
states $N(E)$ of the operator $\tilde{H}_\omega=\tilde{H}+
\tilde{V}_\omega$ which acts on $\ell^2(\Z^{d_1})$. Hence the problem
of surface Lifshitz tails boils down to ordinary Lifshitz tails in a
lower dimensional configuration space. However the (free) operator is
somewhat more complicated, in fact in Fourier representation it is
multiplication by
\begin{equation*}
  \tilde{h}(\theta_1)=\left(\int
  \frac{1}{h(\theta_1,\theta_2)-E_0}\,\; d\theta_2
  \right)^{-1} + E_0.
\end{equation*}
We remind the reader that
$\tilde{V}_\omega(\gamma_1)=\omega_{\gamma_1}$ for
$\gamma_1\in\Z^{d_1}$ and $(\omega_{\gamma_1})_{\gamma_1\in\Z^{d_1}}$
is a family of independent random variables with a common distribution
$P_0$.\\
Throughout this section we assume that supp$(P_0)$ is a compact set.
Moreover, if we set $\omega_-=\inf(\text{supp}(P_0))$ we suppose that
$P_0([\omega_-,\omega_-+\varepsilon)\ge C\; \varepsilon^k)$ for some
$k$.

\begin{Th}\label{thr:5}
  If $\tilde{h}$ has a unique quadratic minimum then
  \begin{equation*}
    \lim_{E\searrow E_0} \frac{\ln |\ln(N_s(E))|}{\ln(E-E_0)} =
    -\frac{d_1}{2}.
  \end{equation*}
\end{Th}
\noindent\textbf{Proof.} The theorem follows
from~\cite{Kl:97b,MR1979772} and the considerations above.\qed
\vskip.2cm\noindent For dimensions $d_1=1$ and $d_1=2$ we have the
following result:
\begin{Th}
  \label{thr:6}
  Assume that $\tilde{h}$ is not constant. If $d_1=1$ then
  \begin{equation*}
    \lim_{E\searrow E_0} \frac{\ln |\ln(N_s(E))|}{\ln(E-E_0)} =
    -\lim_{E\searrow E_0}\frac{\ln(n(E))}{(E-E_0)}
  \end{equation*}
  where $n(E)$ is the integrated density of states for $\tilde{H}$.\\
  If $d_2=2$, then
  \begin{equation*}
    \lim_{E\searrow E_0} \frac{\ln |\ln(N_s(E))|}{\ln(E-E_0)} =
    -\alpha
  \end{equation*}
  where $\alpha$ is defined in~\eqref{eq:86}.
\end{Th}
\noindent Note that $n(E)\sim (E-E_0)^\rho$ for some $\rho>0$.
See~\cite{Kl:97b,MR1979772} for details.
\vskip.2cm\noindent To conclude this section we consider some examples
that fulfill the assumptions of the previous theorems.  Let us first
assume that $H$ is separable, i.e. that
\begin{equation*}
  h(\theta_1,\theta_2)=h_1(\theta_1)+h_2(\theta_2).
\end{equation*}
This is satisfied for example by the discrete Laplacian where $h$ is
equal to $h_0$ given in~\eqref{eq:7}. The function $h$ has a unique
quadratic minimum if and only if both $h_1$ and $h_2$ have unique
quadratic minima (which we may assume to be attained at
$\theta_1=\theta_2=0$).

\noindent We will show in the following that the function

\begin{equation*}
  \tilde{h}(\theta_1)=\left(\int\frac{1}{h_1(\theta_1)+
      h_2(\theta_2)-E_0}\; d\theta_2\right)^{-1} +E_0
\end{equation*}
has a unique quadratic minimum in this case as well. Differentiating
the function
\begin{equation*}
  \rho(\theta_1)=\int \frac{1}{h_1(\theta_1)+h_2(\theta_2)-E_0}\; d\theta_2
\end{equation*}
we obtain:
\begin{equation*}
  \nabla \rho(\theta_1)=-\int \frac{\nabla
    h_1(\theta_1)}{(h_1(\theta_1)+
    h_2(\theta_2)-E_0)^2}\; d\theta_2
\end{equation*}
so the (possible) maximum of $\rho$ is at $\theta_1=0$.

\noindent The second derivative at $\theta_1=0$ is given by:
\begin{equation*}
  \nabla\nabla\rho(0)= - \nabla \nabla h_1(0) \int
  \frac{1}{(h_1(0)+h_2(\theta_2)-E_0)^2}\; d\theta_2
\end{equation*}
which obviously gives a negative definite Hessian.\\
We remark that no assumptions on $h_2$ were needed; in fact, the above
arguments work for $h_2= const$ as well.\\
The same reasoning also shows that $\tilde{h}$ is not constant as long
as $h_1$ is not constant.\\
So we have proved:
\begin{Th}
  \label{thr:7}
  Suppose $h(\theta_1,\theta_2)=h_1(\theta_1)+h_2(\theta_2) $ then
  \begin{enumerate}
  \item If $h_1$ has a unique quadratic minimum, then
    \begin{equation*}
      \lim_{E\searrow E_0} \frac{\ln |\ln(N_s(E))|}{\ln(E-E_0)} =
      -\frac{d_1}{2}.
    \end{equation*}
  \item If $d_1=1$ and $\tilde{h}$ is not constant then
    \begin{equation*}
      \lim_{E\searrow E_0} \frac{\ln |\ln(N_s(E))|}{\ln(E-E_0)} =
      -\lim_{E\searrow E_0}\frac{\ln(n(E))}{(E-E_0)}.
    \end{equation*}
    where $n(E)$ is the integrated density of states for $\tilde{H}$.
  \item If $d_2=2$, then
    \begin{equation*}
    \lim_{E\searrow E_0} \frac{\ln |\ln(N_s(E))|}{\ln(E-E_0)} =
    -\alpha
  \end{equation*}
  where $\alpha$ is defined in~\eqref{eq:86}.
  \end{enumerate}
\end{Th}
\section{The density of surface states for a constant surfacic
  potential}
\label{sec:dens-surf-stat}
In this section, we prove some useful results on the density of
surface states for a constant surface potential. In some cases, this
density may even be computed explicitly (see
e.g.~\cite{MR2000j:81290}).
\par The model we consider is the model introduced in
Proposition~\ref{pro:2} namely $H_t=H+t\car\otimes\Pi_0^2$ where $H$
is chosen as in section~\ref{sec:main-results} and $t$ is a real
coupling constant. The proof of all the results we now state is based
on rank one perturbation theory (see e.g.~\cite{MR97c:47008}). The
main formula that we will use is the following: for $z\not\in\R$, one
has
\begin{equation}
  \label{eq:24}
  (H_t-z)^{-1}-(H-z)^{-1}= -t\frac{1}{1+t I(z)\otimes Id_{\theta_2}}
  (H-z)^{-1}\car\otimes\Pi_0^2(H-z)^{-1}
\end{equation}
where $I(z)$ is the operator acting on $\ell^2(\Z^{d_1})$ that, in
Fourier representation, is the multiplication by the function
$I(\theta_1,z)$ defined in~\eqref{eq:25}.
\par Formula~\eqref{eq:24} is easily proved if one makes a partial Fourier
transform in the $(\gamma_1,\theta_1)$ variable of $H$ and $H_t$. If
one does so, one obtains a direct integral representation for both $H$
and $H_t$ namely%
\begin{equation*}
 H=\int_{\T^{d_1}}H(\theta_1)d\theta_1\text{ and }
 H_t=\int_{\T^{d_1}}H_t(\theta_1)d\theta_1
\end{equation*}
where $H(\theta_1)$ and $H_t(\theta_1)$ (both acting on
$\ell^2(\Z^{d_2})$) differ only by a rank one operator, namely,
\begin{equation*}
  H_t(\theta_1)-H(\theta_1)=t\Pi_0^2.
\end{equation*}
Formulae~\eqref{eq:24} and~\eqref{eq:25} then follow immediately from
the well known resolvent formula for rank one perturbations that can
be found e.g. in~\cite{MR97c:47008}.
\par Proposition~\ref{pro:1} follows immediately from
Proposition~\ref{pro:2} and formulae~\eqref{eq:24} and~\eqref{eq:25}.
Indeed, by formula~\eqref{eq:24} and the special form of the operator
$I(z)$, $z$ is a point in $\sigma(H_t)\setminus\sigma(H)$ if and only
if, for some $\theta_1$, one has
\begin{equation*}
  1+tI(\theta_1,z)=0.
\end{equation*}
If we pick $z\in\R$ below $0$ (recall that $0=\inf(\sigma(H))=
\inf(h(\R^d))$, we see that $z\in\sigma(H_t)$ if and only if
$tI(\theta_1,z)=-1$ for some $\theta_1$. As, for $z<0$,
$I(\theta_1,z)$ is a negative decreasing function of $z$ that tends to
$0$ when $z\to-\infty$, we see that this can happen if an only if
$tI(\theta_1,0)<-1$ for some $\theta_1$. This is the first statement
of Proposition~\ref{pro:4}. Indeed, the function $\theta_1\mapsto
tI(\theta_1,0)$ is continuous of $\T^{d_1}$ except, possibly, at the
points where $h$ assumes its minimum, and it takes its minimal value
exactly at one of those points.
\par As, for the second statement, let $\D
I(z):=\max_{\theta_1\in\T^{d_1}}I(\theta_1,z)$ and consider the
function $f:\ z\mapsto1+tI(z)$. This function is clearly continuous
and strictly decreasing on $]-\infty,0[$ and by assumption, it is
negative near $0$ (as $1+tI_\infty<0$) and $f(z)\to1$ as
$z\to-\infty$. So, the function $f$ admits a unique zero that we
denote by $E_0$. The analysis given above immediately shows that $E_0$
is the infimum of $H_t$: as $\theta_1\mapsto I(\theta_1,z)$ is
continuous on $\T^{d_1}$ that is compact, for some $\theta_1$, one has
$1+tI(\theta_1,E_0)=0$.  So that $E_0$ belongs to $\sigma(H_t)$; on
the other hand, for $E<E_0$, for any $\theta_1$, one has
$1+tI(\theta_1,E)\geq1+tI(E)>0$, hence, $E\not\in\sigma(H_t)$. This
completes the proof of Proposition~\ref{pro:4}.
\subsection{Asymptotics of the density of surface states}
\label{sec:asymptotics}
The starting point for this computation is again
formula~\eqref{eq:24}. This enables us to get a very simple formula
for the Stieltjes-Hilbert transform of the density of surface states
$n_s^t$ for the pair $(H_t,H)$. Using the Fourier representation and
Parseval's formula, one computes
\begin{equation*}
  \begin{split}
    \Tr(\Pi_1[&(H_t-z)^{-1}-(H-z)^{-1}]\Pi_1)\\&=\sum_{\gamma_2\in\Z^{d_2}}
    \int_{\T^{d_1}}\frac{-t}{1+tI(\theta_1,z)}
    \int_{\T^{d_2}}\frac{e^{i\gamma_2\theta_2}d\theta_2}{\overline{h(\theta_1,\theta_2)-z}}
    \int_{\T^{d_2}}\frac{e^{i\gamma_2\theta_2}d\theta_2}{h(\theta_1,\theta_2)-z}d\theta_1\\
    &=\sum_{\gamma_2\in\Z^{d_2}}
    \int_{\T^{d_1}}\frac{-t}{1+tI(\theta_1,z)}\overline{
      \int_{\T^{d_2}}\frac{e^{-i\gamma_2\theta_2}d\theta_2}{h(\theta_1,\theta_2)-z}}
    \int_{\T^{d_2}}\frac{e^{i\gamma_2\theta_2}d\theta_2}{h(\theta_1,\theta_2)-z}d\theta_1\\
    &=\int_{\T^{d_1}}\frac{-t}{1+tI(\theta_1,z)}
    \int_{\T^{d_2}}\frac{d\theta_2}{(h(\theta_1,\theta_2)-z)^2}d\theta_1
  \end{split}
\end{equation*}
One then notices that
\begin{equation*}
  \int_{\T^{d_1}}\frac{-t}{1+tI(\theta_1,z)}
  \int_{\T^{d_2}}\frac{d\theta_2}{(h(\theta_1,\theta_2)-z)^2}d\theta_1=
  -\frac{d}{dz}\int_{\T^{d_1}}\log(1+tI(\theta_1,z))d\theta_1.
\end{equation*}
Here, and in the sequel, $\log$ denotes the principal determination of
the logarithm.\\
This immediately yields that the Stieltjes-Hilbert transform of
$N_s^t$ is given by
\begin{equation*}
  \left\langle\frac{1}{\cdot-z},dN_s^t\right\rangle=
  \int_{\T^{d_1}}\log(1+tI(\theta_1,z))d\theta_1
\end{equation*}
where $I$ is defined by~\eqref{eq:25}.\\
It is well known that one can invert the Stieltjes-Hilbert transform
to recover the signed measure $dN_s^t$ (see e.g. the appendix
of~\cite{MR94h:47068}). By the Stieltjes-Perron inversion formula, one
has
\begin{equation}
  \label{eq:28}
  \begin{split}
    \int_0^EdN_s^t(e)&=\lim_{\varepsilon\to0^+}\frac{1}{2i\pi}\int_0^E
    \left(\left\langle\frac{1}{\cdot-e-i\varepsilon},dN_s^t\right\rangle-
      \left\langle\frac{1}{\cdot-e+i\varepsilon},dN_s^t\right\rangle\right)de\\
    &=\lim_{\varepsilon\to0^+}\frac{1}{2i\pi}\int_0^E\int_{\T^{d_1}}
    [\log(1+tI(\theta_1,e+i\varepsilon))-\log(1+tI(\theta_1,e-i\varepsilon))]
    d\theta_1de.
  \end{split}
\end{equation}
Notice that, for $e$ real,
\begin{equation*}
  \text{Im}\left(1+tI(\theta_1,e+i\varepsilon)\right)=
  t\varepsilon
  \int_{\T^{d_2}}\frac{1}{(h(\theta_1,\theta_2)-e)^2+\varepsilon^2}d\theta_2;
\end{equation*}
hence, this imaginary part keeps a fixed sign. So, for
$\theta_1\in\T^{d_1}$, one has
\begin{equation*}
  \log(1+tI(\theta_1,e+i\varepsilon))-\log(1+tI(\theta_1,e-i\varepsilon))
  =\log
  \left(\frac{1+tI(\theta_1,e+i\varepsilon)}{1+tI(\theta_1,e-i\varepsilon)}\right)
\end{equation*}
For $e\in\R$, one has $|1+tI(\theta_1,e+i\varepsilon)|=
|1+tI(\theta_1,e-i\varepsilon)|$. As moreover the imaginary part of
$1+tI(\theta_1,e+i\varepsilon)$ keeps a fixed sign, one has
\begin{equation*}
  |\log(1+tI(\theta_1,e+i\varepsilon))-\log(1+tI(\theta_1,e-i\varepsilon))|\leq 2\pi.
\end{equation*}
As $\T^{d_1}$ and $[0,E]$ are compact, one can apply Lebesgue's
dominated convergence Theorem to~\eqref{eq:28} and thus obtain
\begin{equation}
  \label{eq:30}
  \int_0^EdN_s^t(e)=\int_0^E\int_{\T^{d_1}}f(\theta_1,e) d\theta_1 de.
\end{equation}
where
\begin{equation}
  \label{eq:31}
  f(\theta_1,e)=\lim_{\varepsilon\to0^+}\frac{1}{2i\pi}\log
  \left(\frac{1+tI(\theta_1,e+i\varepsilon)}{1+tI(\theta_1,e-i\varepsilon)}\right)
  =\lim_{\varepsilon\to0^+}\frac{1}{\pi}\text{Arg}(1+tI(\theta_1,e+i\varepsilon))
\end{equation}
where Arg is the principal determination of the argument of a complex
number. Notice here that this formula is the analogue of the
well-known Birman-Kre\u\i n formula (see
e.g.~\cite{MR94g:47002,Ya:92}) for surface perturbations.
\par We will now compute the asymptotics of $f(\theta_1,e)$ for $e$
small. First, let us notice that we need only to compute these for
$\theta_1$ small, i.e. close to $0$. Indeed, we have assumed that $h$
takes its minimum only at $0$. Therefore, as $\T^d$, is compact, if
$|\theta_1|\geq\delta$, we know that, for some $\delta'>0$, for all
$\theta_2$, one has $h(\theta_1,\theta_2)\geq\delta'$. Hence, if
$|\theta_1|\geq\delta$, the function $I(\theta_1,z)$ is analytic in a
neighborhood of $0$, so that $f(\theta_1,e)=0$ for $e$ sufficiently
small (independent of $\theta_1$). So, we now assume that
$|\theta_1|<\delta$ for some $\delta>0$ to be chosen later on.
\par We now study $I(\theta_1,z)$ for $|z|$ small. Pick $\chi$ a
smooth cut-off function in $\theta_2$, i.e. such that
$\chi(\theta_2)=1$ if $|\theta_2|\leq\delta_\chi$ and
$\chi(\theta_2)=0$ if $|\theta_2|\geq2\delta_\chi$. Write
\begin{equation}
  \label{eq:39}
  I(\theta_1,z)=
  \int_{\T^{d_2}}\frac{\chi(\theta_2)}{h(\theta_1,\theta_2)-z}d\theta_2+
  \int_{\T^{d_2}}\frac{1-\chi(\theta_2)}{h(\theta_1,\theta_2)-z}d\theta_2.
\end{equation}
For the same reason as above, the second integral in the right hand
side term is analytic for $|z|$ small for all $\theta_1$. We only need
to study the integral
\begin{equation}
  \label{eq:33}
  J(\theta_1,z)=
  \int_{\T^{d_2}}\frac{\chi(\theta_2)}{h(\theta_1,\theta_2)-z}d\theta_2.
\end{equation}
Therefore, we use the assumptions that $0$ is the unique minimum of
$h$ and that it is quadratic non-degenerate. This implies, that for
$\delta>0$ sufficiently small, for $|\theta_1|<\delta$, the function
$\theta_2\mapsto h(\theta_1,\theta_2)$ has a unique minimum, say
$\theta_2(\theta_1)$, that this minimum is quadratic non-degenerate.
Let $h_2(\theta_1)$ be the minimal value, i.e.
$h_2(\theta_1)=h(\theta_1,\theta_2(\theta_1))$. Then, the functions
$\theta_1\mapsto\theta_2(\theta_1)$ and $\theta_1\mapsto h(\theta_1)$
are real analytic in $|\theta_1|<\delta$.\\
All these statements are immediate consequences of the analytic
Implicit Function Theorem applied to the system of equations
$\nabla_{\theta_2}h(\theta_1,\theta_2)=0$.\\
So, for $|\theta|<\delta$, one can write
\begin{equation*}
  h(\theta_1,\theta_2)=h_2(\theta_1)+\langle(\theta_2-\theta_2(\theta_1))
  ,Q_2(\theta_1)(\theta_2-\theta_2(\theta_1))\rangle+O(|\theta_2-\theta_2(\theta_1)|^3)
\end{equation*}
where $Q_2(\theta_1)$ is the Hessian matrix of $h(\theta_1,\theta_2)$
at
the point $\theta_2(\theta_1)$.\\
We can now use the analytic Morse Lemma (see e.g.~\cite{Hor:90})
uniformly in the parameter $\theta_1$. That is, for some $\delta_0>0$
small, there exists $B_2(0,\delta_0)\subset U$ (the ball of center $0$
and radius $\delta_0$ in $\T^{d_2}$) and $\psi(\theta_1):\ \theta_2\in
U\to \psi(\theta_1,\theta_2)\in B_2(\theta_2(\theta_1),2\delta_0)$, a
real analytic diffeomorphism so that, for $\theta\in U$,
\begin{equation}
  h(\theta_1,\psi(\theta_1,\theta_2))=
  h_2(\theta_1)+(\theta_2,Q_2(\theta_1)\theta_2).
\end{equation}
Moreover, the Jacobian matrix of $\psi$ at $\theta_2(\theta_1)$ is the
identity matrix, and the mapping $\theta_1\mapsto\psi(\theta_1)$ is
real analytic (here, we take the norm in the Banach space of real
analytic function in a neighborhood of $0$).\\
Before we return to the analysis of $J$, let us describe $h_2(\cdot)$
and $\theta_2(\cdot)$ more precisely. Let $Q$ be the Hessian matrix of
$h$ at $0$. As $h$ has a quadratic non degenerate minimum at $0$, $Q$
is definite positive. We can write this $d\times d$-matrix in the form
\begin{equation}
  \label{eq:36}
  Q=\begin{pmatrix} Q_1 & R^*\\ R & Q_2 \end{pmatrix}
\end{equation}
where $Q_{1,2}$ is the restriction of $Q$ to $\R^{d_1,d_2}$ when one
decomposes $\R^d=\R^{d_1}\times\R^{d_2}$. Both $Q_1$ and $Q_2$ are
positive definite; actually, the positive definiteness of $Q$ ensures
that the matrices $Q_1-R^*Q_2^{-1}R$ and $Q_2-RQ_1^{-1}R^*$ are
positive definite. Using the Taylor expansion of $h$ near $0$, one
computes
\begin{gather}
  \nonumber\theta_2(\theta_1)=-Q_2^{-1}R\theta_1+O(|\theta_1|^2),\quad
  Q_2(\theta_1)=Q_2+O(|\theta_1|),\\
  \label{eq:45}
  h_2(\theta_1)=([Q_1-R^*Q_2^{-1}R]\theta_1,\theta_1)+O(|\theta_1|^3).
\end{gather}
Let us also note here that
\begin{equation}
  \label{eq:46}
  \text{Det}\,Q=\text{Det}\,Q_1\cdot\text{Det}\,(Q_2-R^*Q_1^{-1}R)=
  \text{Det}\,Q_2\cdot\text{Det}\,(Q_1-RQ_2^{-1}R^*)
\end{equation}
We now return to $J$. Performing the change of variables
$\theta\to\psi(\theta)$ in $J(\theta_1,z)$, we get
\begin{equation}
  \label{eq:38}
    J(\theta_1,z)=
  \int_{\T^{d_2}}\frac{\tilde\chi(\theta_1,\theta_2)}
  {(\theta_2,Q_2(\theta_1)\theta_2)+h_2(\theta_1)-z}d\theta_2\text{
    where } \tilde\chi(\theta_1,\theta_2):=
  \chi(\psi(\theta_1,\theta_2))\text{Det}\,(
    \nabla_{\theta_2}\psi(\theta_1,\theta_2)).
\end{equation}
Choosing $\delta$ sufficiently small with respect to $\delta_\chi$
(defining $\chi$), we see that $\chi(\psi(\theta_1,\theta_2))=1$ for
all $|\theta_1|<\delta$ and $|\theta_2|<\delta$.  Hence, the function
$\tilde\chi(\theta_1,\theta_2)$ is real analytic in a neighborhood of
$(0,0)$.\par
To compute the integral in the right hand side of~\eqref{eq:38}, we
change to polar coordinates (recall that $\tilde\chi$ is supported
near $0$) to obtain
\begin{equation}
  \label{eq:35}
  J(\theta_1,z)=\text{Det}\,(Q_2(\theta_1))^{-1/2}
  \int_0^{+\infty}\frac{\hat\chi(\theta_1,r)r^{d_2-1}}{r^2+h_2(\theta_1)-z}dr
\end{equation}
where
\begin{equation}
  \label{eq:64}
  \check\chi(\theta_1,r):=\frac{1}{(2\pi)^{d_2}}\int_{\mathbb{S}^{d_2-1}}
  \tilde\chi(\theta_1,r\xi)d\xi.
\end{equation}
The factor $(2\pi)^{-d_2}$ in the last integral comes from the fact
that $d\theta_2$ denotes the normalized Haar measure on $\T^{d_2}$,
i.e. the Lebesgue measure divided by $(2\pi)^{d_2}$. Note again that
$(\theta_1,r)\mapsto\check\chi(\theta_1,r)$ is real analytic in a
neighborhood of $0$, and
\begin{equation*}
  \check\chi(\theta_1,0)=\frac{1}{(2\pi)^{d_2}}\text{Det}\,(\nabla_{\theta_2}
  \psi(\theta_1,\theta_2(\theta_1)))\cdot\text{Vol}(\mathbb{S}^{d_2-1}).
\end{equation*}
Moreover, as $\int_{\mathbb{S}^{d_2}}\xi^kd\xi=0$ if $k$ is
multi-index of odd length, we known that the Taylor expansion of
$\check\chi(\theta_1,r)$ contains only even powers of $r$, i.e. there
exists a function $\hat\chi(\theta_1,r)$ analytic in a neighborhood of
$(0,0)$ such that $\check\chi(\theta_1,r)=\hat\chi(\theta_1,r^2)$.
\par We now use the
\begin{Le}
  \label{le:2}
  Let $\hat\chi$ be a smooth compactly supported function such that
  $\hat\chi$ be real analytic is a neighborhood of $0$. Define the
  integral $J_{\hat\chi}(z)$ to be
  \begin{equation*}
    J_{\hat\chi}(z)=\int_0^{+\infty}\frac{\hat\chi(r^2)r^{n-1}}{r^2+z}dr.
  \end{equation*}
  Then, one has
  \begin{equation}
    \label{eq:37}
    J_{\hat\chi}(z)=S(z)\cdot H(z) + G(z)
  \end{equation}
  where
  \begin{enumerate}
  \item $G$ and $H$ are real analytic in a neighborhood of $0$;
  \item they satisfy $H(0)=\hat\chi(0)$ and $G(0)>0$ if
    $\hat\chi(0)>0$ and $\hat\chi\geq0$;
  \item the function $S$ is defined by
    \begin{itemize}
    \item if $n$ is even, then $\D S(z)=\frac{1}{2}\cdot
      (-1)^{\frac{n}{2}}z^{\frac{n-2}{2}}\cdot\log z$;
    \item if n is odd, then $\D S(z)=\frac{\pi}{2}\cdot
      (-1)^{\frac{n-1}{2}}z^{\frac{n-1}{2}}\frac{1}{\sqrt{z}}$.
    \end{itemize}
    Here, $\sqrt{z}$ and $\log z$ denote respectively the principal
    determination of the square root and of the logarithm.
\end{enumerate}
\end{Le}
\noindent The proof of this result is elementary; after a cut-off near
zero, one expands $\hat\chi$ in a Taylor series near $0$, and computes
the resulting integrals term by term essentially explicitly (see
~\cite{Kl:94a} for more details).
\par
Putting~\eqref{eq:39},~\eqref{eq:33},~\eqref{eq:38},~\eqref{eq:35}
and~\eqref{eq:37} together, we obtain that
\begin{equation}
  \label{eq:40}
  I(\theta_1,z)=S(h_2(\theta_1)-z)\cdot H(\theta_1,h_2(\theta_1)-z) +
  G(\theta_1,h_2(\theta_1)-z)
\end{equation}
where
\begin{itemize}
\item $S$ is described in point (3) of Lemma~\ref{le:2};
\item $(\theta_1,z)\mapsto H(\theta_1,z)$ and $(\theta_1,z)\mapsto
  G(\theta_1,z)$ are real analytic in $\theta_1$ and $z$ in a
  neighborhood of $0$;
\item one has
  \begin{equation*}
    H(\theta_1,0)=\frac{1}{(2\pi)^{d_2}}
    \text{Det}\,(Q_2(\theta_1))^{-1/2}\cdot
    \text{Det}\,(\nabla_{\theta_2}\psi(\theta_1,\theta_2))
    \text{Vol}(\mathbb{S}^{d_2-1})
  \end{equation*}
  and $G(0,0)$ is positive.
\end{itemize}
The last point here is obtained combining point (2) of
Lemma~\ref{le:2},~\eqref{eq:35} and~\eqref{eq:64}, and using the
decomposition~\eqref{eq:39}.
\par The first immediate consequence of~\eqref{eq:40} is that, if
$e\in\R$ and $h_2(\theta_1)> e$, then
\begin{equation*}
 I(\theta_1,e+i\varepsilon)-I(\theta_1,e-i\varepsilon)\to0\text{ when
 }\varepsilon\to0^+.
\end{equation*}
This implies that, if $h_2(\theta_1)> e$, one has
\begin{equation*}
  f(\theta_1,e)=0.
\end{equation*}
Assume now that $h_2(\theta_1)\leq e$. As $0\leq h_2(\theta_1)$,
$-e\leq h_2(\theta_1)-e\leq 0$. We now need to distinguish different
cases according to the dimension $d_2$. Consider the case
\begin{itemize}
\item $d_2=1$: by~\eqref{eq:40}, as $H$ and $G$ are analytic, one has
  \begin{equation*}
    \lim_{\varepsilon\to0^+}I(\theta_1,e+i\varepsilon)=-\frac{\pi}{2}
    \frac{i}{\sqrt{|h_2(\theta_1)-e|}}H(\theta_1,h_2(\theta_1)-e)
    +G(\theta_1,h_2(\theta_1)-e).
  \end{equation*}
  Using again the fact that $H$ and $G$ are analytic and that
  $H(\theta_1,0)$ does not vanish for $\theta_1$ small, we get
  \begin{equation*}
    \lim_{\varepsilon\to0^+}\frac{1+tI(\theta_1,e+i\varepsilon)}
    {1+tI(\theta_1,e-i\varepsilon)}=-1+i\frac{2G(0,0)\sqrt{|h_2(\theta_1)-e|}}{tH(0,0)}
  +o\left(\sqrt{|h_2(\theta_1)-e|}\right).
  \end{equation*}
  As $G(0,0)$, $H(0,0)$ and $t$ are also positive, one finally obtains
  \begin{equation*}
    f(\theta_1,e)=\frac{1}{2}\left[1+O\left(\sqrt{|h_2(\theta_1)-e}|\right)
    \right]\cdot\car_{\{h_2(\theta_1)\leq e\}}.
  \end{equation*}
\item $d_2=2$: in this case, one computes
  \begin{equation*}
    \lim_{\varepsilon\to0^+}I(\theta_1,e+i\varepsilon)=\frac{1}{2}
    (|\log|h_2(\theta_1)-e||+i\pi)H(\theta_1,h_2(\theta_1)-e)
    +G(\theta_1,h_2(\theta_1)-e).
  \end{equation*}
  Using again the fact that $H$ and $G$ are analytic, we get
  \begin{equation*}
    \lim_{\varepsilon\to0^+}\frac{1+tI(\theta_1,e+i\varepsilon)}
    {1+tI(\theta_1,e-i\varepsilon)}=\left(1+
      \frac{2i\pi}{|\log|h_2(\theta_1)-e||}\right)\cdot
    (1+O[(\log|h_2(\theta_1)-e|)^{-1}]).
  \end{equation*}
  So that finally, one has
  \begin{equation*}
    f(\theta_1,e)=
    \frac{1}{|\log|h_2(\theta_1)-e||}(1+O[(\log|h_2(\theta_1)-e|)^{-1}])
    \cdot\car_{\{h_2(\theta_1)\leq e\}}.
  \end{equation*}

\item $d_2\geq3$: in this case, one has to distinguish two cases
  whether $1+tI(0,0)=0$ or not, as well as the case of even and odd
  dimensions.\\
  Let us first assume:
  \begin{itemize}
  \item that $1+tI(0,0)>0$: as $H$ and $G$ are analytic, one has
    \begin{equation*}
      \lim_{\varepsilon\to0^+}I(\theta_1,e+i\varepsilon)=\left(
        \lim_{\varepsilon\to0^+}S(h_2(\theta_1)-e-i\varepsilon)\right)
      \cdot H(\theta_1,h_2(\theta_1)-e)+G(\theta_1,h_2(\theta_1)-e).
    \end{equation*}
    As $G$ is analytic and as $S(0)=0$, one has $G(\theta_1,
    h_2(\theta_1))=I(\theta_1,0)$. So, for $\theta_1$ small, we know
    that $1+tG(\theta_1,0)\not=0$. Here, we used the continuity of $G$
    and the fact that $h_2(\theta_1)$ is of size $|\theta_1|^2$ hence
    small. This gives
    \begin{equation*}
      \lim_{\varepsilon\to0^+}\frac{1+tI(\theta_1,e+i\varepsilon)}
      {1+tI(\theta_1,e-i\varepsilon)}=1+\frac{t\cdot s(h_2(\theta_1)-e)
        \cdot H(\theta_1,0)}{1+t\cdot G(\theta_1,0)}\cdot(1+R)
    \end{equation*}
    where
    \begin{equation*}
      s(x)=\lim_{\varepsilon\to0^+}[S(x-i\varepsilon)-S(x+i\varepsilon)],\\
      \nonumber
      R=O((h_2(\theta_1)-e)\cdot|S(h_2(\theta_1)-e)|,(h_2(\theta_1)-e)).
    \end{equation*}
    So that finally, for $e$ small, one has
    \begin{equation}
      f(\theta_1,e)=\frac{t\cdot s(h_2(\theta_1)-e)\cdot H(\theta_1,0)}
      {1+tG(\theta_1,0)}\car_{\{h_2(\theta_1)\leq e\}}(1+R).
    \end{equation}
    where
    \begin{equation}
      \label{eq:50}
      s(x)=\frac{1}{2}|x|^{\frac{d_2-2}{2}}
    \end{equation}
    and $R$ is given above.
  \end{itemize}
\end{itemize}
From these asymptotics and from~\eqref{eq:30}, integrating $f$
in~\eqref{eq:30}, using~\eqref{eq:45} and~\eqref{eq:46}, one gets that
\begin{itemize}
\item if $d_2=1$:
  \begin{equation}
    \label{eq:55}
  \int_0^EdN_s^t(e)\equ_{E\to0^+}
  \D\frac{\text{Vol}(\mathbb{S}^{d_1-1})}
  {d_1(d_1+2)(2\pi)^{d_1}\sqrt{\text{Det}\,(Q_1-RQ_2^{-1}R^*)}} \cdot
  E^{1+d_1/2}
  \end{equation}
\item if $d_2=2$:
  \begin{equation*}
  \int_0^EdN_s^t(e)\equ_{E\to0^+} \frac{2\text{Vol}(\mathbb{S}^{d_1-1})}
  {d_1(d_1+2)(2\pi)^{d_1}\sqrt{\text{Det}\,(Q_1-RQ_2^{-1}R^*)}}
  \frac{E^{1+d_1/2}}{|\log E|}
  \end{equation*}
\item if $d_2\geq3$ and $1+tI(0,0)>0$:
  \begin{equation}
    \label{eq:57}
  \int_0^EdN_s^t(e)\equ_{E\to0^+}
  \frac{c(d_1,d_2)\text{Vol}(\mathbb{S}^{d_2-1})\text{Vol}(\mathbb{S}^{d_1-1})
    }{d(2\pi)^{d}\sqrt{\text{Det}\,Q}}\cdot \frac{t}{1+tI(0,0)}\cdot s(E)E^{1+d_1/2}
  \end{equation}
\end{itemize}
where $\mathbb{S}^{d_{1,2}-1}$ are respectively the $d_{1,2}-1$
dimensional unit spheres, and $s$ is given by~\eqref{eq:50}. Here,
$c(d_1,d_2)$ is the integral
\begin{equation}
  \label{eq:66}
  c(d_1,d_2)=\int_0^1r^{d_1-1}(1-r^2)^{(d_2-2)/2}dr.
\end{equation}
\subsubsection{The borderline case}
\label{sec:borderline-case}
Though it will not find direct applications in this paper, let us now
turn to the case when $d_2\geq3$ and $1+tI(0,0)=0$. Notice that this
assumption implies $t<0$. When $1+tI(0,0)=0$, one has to take a closer
look at the vanishing of $1+tG(\theta_1,0)$ when $\theta_1\to0$. We
will now assume that
\begin{description}
\item[(H)] $I(\theta_1,0)$ has a local maximum for $\theta_1=0$.
\end{description}
\begin{Rem}
  \label{rem:3}
  Notice that this assumption was also necessary for fluctuating
  edges. Actually, in that setting, we even required that the maximum
  be non-degenerate if $d_1\geq3$. This seems quite natural as the
  case $1+tI(0,0)=0$ is exactly the border line between the
  fluctuating edges and stable edges.
\end{Rem}
Let us recall that as above, we need to compute the asymptotic when
$e\to0^+$ of the integral
\begin{equation*}
  \int_{\T^{d_1}}f(\theta_1,e) d\theta_1=
  \int_{\{h_2(\theta_1)\leq e\}}f(\theta_1,e)d\theta_1
\end{equation*}
where $f$ is defined by~\eqref{eq:31}. Using~\eqref{eq:45} we can find
a analytic change of variable $\theta_1\mapsto\psi(\theta_1)$ such
that $h_2(\psi^{-1}(\theta_1))=\langle\tilde
Q_1\theta_1,\theta_1\rangle=:q_2(\theta_1)$ where $\tilde
Q_1=Q_1-R^*Q_2^{-1}R$ (the matrices $Q_{1,2}$ and $R$ are defined
in~\eqref{eq:36}) and $\psi(\theta_1)=\theta_1+O(|\theta_1|^2)$. So,
we want to study
\begin{equation*}
  \int_{\{q_2(\theta_1)\leq e\}}f(\psi(\theta_1),e)
  |\text{Det}\nabla_{\theta_1}\psi(\theta_1)|d\theta_1
\end{equation*}
Let us perform one more change of variable in the integral above,
namely $\theta_1\leftrightarrow\sqrt{e}\theta_1$; hence, we need to
study
\begin{equation*}
  \int_{\{q_2(\theta_1)\leq 1\}}f(\psi(\sqrt{e}\theta_1),e)
  |\text{Det}\nabla_{\theta_1}\psi(\sqrt{e}\theta_1)|d\theta_1
\end{equation*}
Notice that, for $e$ small, on $\{\langle\tilde
Q_1\theta_1,\theta_1\rangle\leq 1\}$, one has
\begin{equation*}
  |\text{Det}\nabla_{\theta_1}\psi(\sqrt{e}\theta_1)|
  =1+O(\sqrt{e}).
\end{equation*}
We now study $f(\psi(\sqrt{e}\theta_1),e)$ for $e$ small and
$\{q_2(\theta_1)\leq 1\}$.\\
Using the analyticity of $G$ and $H$, for $\varepsilon>0$, we start
with rewriting~\eqref{eq:40} in the following way
\begin{multline}
  \label{eq:58}
  1+tI(\psi(\sqrt{e}\theta_1),e+i\varepsilon)=
  1+tG(\psi(\sqrt{e}\theta_1),0)+
  te\partial_z G(0,0)(q_2(\theta_1)-1)+\\+
  tS(e\cdot(q_2(\theta_1)-1)-i\varepsilon))
  H(\psi(\sqrt{e}\theta_1),0)
  +O(\varepsilon+e^2+|e\cdot S(e)|).
\end{multline}
Let us now distinguish between the different dimensions, i.e. between
the cases $d_2=3$, $d_2=4$ and $d_2\geq5$. Substituting the
asymptotics for $S$ given in Lemma~\ref{le:2} and using the
analyticity of $G$ and $H$, one obtains the following:
\begin{itemize}
\item If $d_2=3$: define $\D F_\pm(\theta_1,e)=
  \lim_{\varepsilon\to0^+} 1+tI(\psi(\sqrt{e}\theta_1),e\pm
  i\varepsilon)$. For $q_2(\theta_1)<1$, one has
  \begin{equation*}
    F_\pm(\theta_1,e)=\sqrt{e}\left(\mp
    it(2\pi)^{-d_2}|1-q_2(\theta_1)|^{1/2}
    \text{Det}\,(Q_2)^{-1/2}+t\cdot g(\theta_1)+o(\sqrt{e})\right)
  \end{equation*}
  where
  \begin{equation}
    \label{eq:51}
    t\cdot g(\theta_1)=\lim_{e\to0^+}\frac{1}{\sqrt{e}}
    [1+tG(\psi(\sqrt{e}\theta_1),0)].
  \end{equation}
  This gives, for $q_2(\theta_1)<1$,
  \begin{equation*}
    f(\psi(\sqrt{e}\theta_1),e)\equ_{e\to0^+}
    \frac{1}{\pi}\text{Arg}(-i(2\pi)^{-d_2}\text{Det}\,(Q_2)^{-1/2}
    |1-q_2(\theta_1)|^{1/2}+g(\theta_1)).
  \end{equation*}
  We notice that this last argument is non positive. As a result we
  obtain that
  \begin{equation}
    \label{eq:54}
    \int_0^EdN_s^t(e)de\equ_{E\to0^+}
    \D\frac{\D\int_{|\theta_1|\leq1}\text{Arg}(-i
      |1-\theta_1^2|^{1/2} +\tilde
    g(\theta_1))d\theta_1}{d_1(d_1+2)\pi(2\pi)^{d_1}
    \sqrt{\text{Det}\,(Q_1-RQ_2^{-1}R^*)}} \cdot E^{1+d_1/2}
  \end{equation}
  where
  \begin{equation*}
    \tilde g(\theta_1)=(2\pi)^{d_2}\sqrt{\text{Det}\,(Q_2)}
    g((Q_1-RQ_2^{-1}R^*)^{-1/2}\theta_1)
  \end{equation*}
  and $g$ is defined by~\eqref{eq:51}.
  \begin{Rem}
    \label{rem:2}
    In some cases, $\tilde g$ and $g$ are identically vanishing. This
    happens for example if $h$ is a ``separate variable'' function,
    i.e. if $h(\theta_1,\theta_2)=\tilde h_1(\theta_1)+ \tilde
    h_2(\theta_2)$.  Indeed, in this case, $h_2(\theta_1)=\tilde
    h_1(\theta_1)$ and $I(\theta_1,h_2(\theta_1))=I(0,0)$, hence, $G$
    does not depend on $\theta_1$, i.e. $G(\theta_1,z)=G(z)$.\\
    When $\tilde g$ vanishes identically, formula~\eqref{eq:54}
    becomes~\eqref{eq:55} except for the sign which changes to $-$.\\
    The integral $\D\int_{|\theta_1|\leq1}\text{Arg}(-it|1-\theta_1^2
    |^{1/2} +\tilde g(\theta_1))d\theta_1$ is negative.  Hence,
    comparing~\eqref{eq:54} to~\eqref{eq:57}, we see that,
    asymptotically when $E\to0^+$, $\int_0^EdN_s^t(e)de$ is larger
    when $1+tI(0,0)=0$ than when $1+tI(0,0)>0$. This is explained by
    the fact that, when $1+tI(0,0)=0$, a zero energy resonance (or
    eigenvalue if $d_2\geq5$) is created. This resonance (eigenvalue)
    carries more weight. Of course, the same phenomenon
    happens for the spectral shift function.\\
    To conclude the case $d_2=3$, let us notice that we did not use
    assumption (H).
  \end{Rem}
\item If $d_2=4$: let us start with computing
  $\partial_{\theta_1}G(0,0)$. Therefore, we use
  $G(\theta_1,0)=I(\theta_1,h_2(\theta_1))$ and compute
  \begin{equation*}
    \begin{split}
      \partial_{\theta_1}G(0,0)&=\partial_{\theta_1}
      [I(\theta_1,h_2(\theta_1))]_{|\theta_1=0}=
      -\left(\int_{\T^{d_2}}\frac{\partial_{\theta_1}
          (h(\theta_1,\theta_2)-h_2(\theta_1))}
        {(h(\theta_1,\theta_2)-h_2(\theta_1))^2}d\theta_1
      \right)_{|\theta_1=0}\\&=-\int_{\T^{d_2}}
      \frac{\partial_{\theta_1}h(0,\theta_2)}
      {(h(0,\theta_2))^2}d\theta_1=0
    \end{split}
  \end{equation*}
  as $0$ is a local maximum of $I(\theta_1,0)$. This computation
  immediately gives that $1+tG(\psi(\sqrt{e}\theta_1),0)=O(e)$. Hence,
  equation~\eqref{eq:58} gives

  \begin{equation*}
    F_\pm(\theta_1,e)=te(q_2(\theta_1)-1)(\log e+
    \log|q_2(\theta_1)-1|+h(e))\cdot\left(1+\frac{\mp
        i\pi}{\log e+\log|q_2(\theta_1)-1|+h(e)}\right)
  \end{equation*}
  where $h(e)$ is bounded and does not depend on the sign $\pm$. This
  gives, for $q_2(\theta_1)<1$,
  \begin{equation*}
    f(\psi(\sqrt{e}\theta_1),e)\equ_{e\to0^+}-\frac{1}{|\log e|}.
  \end{equation*}
  Integrating over $\theta_1$ and $e$, we obtain
  \begin{equation*}
  \int_0^EdN_s^t(e)\equ_{E\to0^+} -\frac{2\text{Vol}(\mathbb{S}^{d_1-1})}
  {d_1(d_1+2)(2\pi)^{d_1}\sqrt{\text{Det}\,(Q_1-RQ_2^{-1}R^*)}}
  \frac{E^{1+d_1/2}}{|\log E|}.
  \end{equation*}
\item If $1+tI(0,0)=0$ and $d_2\geq5$: we now compute
  $\partial^2_{\theta_1}Q(0,0)$. Therefore, we continue the
  computation done above to obtain
  \begin{equation}
    \label{eq:62}
    \begin{split}
      \partial^2_{\theta_1}G(0,0)&=-\partial_{\theta_1}
      \left(\int_{\T^{d_2}}\frac{\partial_{\theta_1}
          (h(\theta_1,\theta_2)-h_2(\theta_1))}
        {(h(\theta_1,\theta_2)-h_2(\theta_1))^2}d\theta_1
      \right)_{|\theta_1=0}\\&=
      -\left(\int_{\T^{d_2}}\frac{\partial^2_{\theta_1}
          (h(\theta_1,\theta_2)-h_2(\theta_1))}
        {(h(\theta_1,\theta_2)-h_2(\theta_1))^2}d\theta_1
      \right)_{|\theta_1=0}+2
      \left(\int_{\T^{d_2}}\frac{[\partial_{\theta_1}
          (h(\theta_1,\theta_2)-h_2(\theta_1))]^2}
        {(h(\theta_1,\theta_2)-h_2(\theta_1))^3}d\theta_1
      \right)_{|\theta_1=0}\\&=
      -\int_{\T^{d_2}}\frac{\partial^2_{\theta_1}
        h(0,\theta_2)}{(h(0,\theta_2))^2}d\theta_1
      +2\int_{\T^{d_2}}\frac{[\partial_{\theta_1}
        h(0,\theta_2)]^2}{(h(0,\theta_2))^3}d\theta_1
      +\left(\int_{\T^{d_2}}\frac{1}{(h(0,\theta_2))^2}d\theta_1
      \right)Q_2
    \end{split}
  \end{equation}
  where $Q_2$ is defined in~\eqref{eq:36}. On the other hand, one has
  \begin{equation*}
    \partial_z G(0,0)=-\partial_z I(0,z)_{|z=0}=-J\quad\text{ where
    }\quad J:=\int_{\T^{d_2}}\frac{1}{(h(0,\theta_2))^2}d\theta_1.
  \end{equation*}
  Plugging this and~\eqref{eq:62} into~\eqref{eq:58}, we obtain
  \begin{equation*}
    1+tI(\psi(\sqrt{e}\theta_1),e\pm i\varepsilon)=-teJ+o(e)
  +tS(e\cdot(q_2(\theta_1)-1)\mp i\varepsilon)(H(0,0)+o(1)).
  \end{equation*}
  where $o(e)$ does not depend of $\pm$. This gives, for
  $q_2(\theta_1)<1$,
  \begin{equation*}
    f(\psi(\sqrt{e}\theta_1),e)\equ_{e\to0^+}
    -\frac{s(e\cdot(q_2(\theta_1)-1))}{J}.
  \end{equation*}
  Integrating over $\theta_1$ and $e$, we obtain
  \begin{equation*}
    \int_0^EdN_s^t(e)\equ_{E\to0^+}
    \frac{c(d_1,d_2)\text{Vol}(\mathbb{S}^{d_2-1})\text{Vol}(\mathbb{S}^{d_1-1})
    }{d(2\pi)^{d}\sqrt{\text{Det}\,Q}}\cdot \frac{-1}{J}\cdot s(E)E^{d_1/2}
  \end{equation*}
  where $c(d_1,d_2)$ is defined in~\eqref{eq:66}.
\end{itemize}
\section{Appendix}
\label{sec:appendix}
Pick $E<-d$. We now prove that, for $h$ taken as in
Remark~\ref{rem:1}, the function $\tilde h$ defined in~\eqref{eq:3} is
not constant. For the purpose of this argument, we write
$\theta_1=(\theta^1,\cdots,\theta^{d_1})$. To check that $\tilde h $
is not constant, by~\eqref{eq:25} and~\eqref{eq:3}, it suffices to
check that the function $\theta_1\mapsto I(\theta_1,E)$ is not
constant, hence, that the function $\theta^1\mapsto J(\theta^1)$
defined by
\begin{equation}
  \label{eq:4}
  J(\theta^1)=\frac{1}{(2\pi)^{d_1-1}}
  \int_{[0,2\pi]^{d_1-1}}I(\theta^1,\theta^2,\cdots,\theta^{d_1},E)
  d\theta^2\cdots d\theta^{d_1}=\frac{1}{(2\pi)^{d-1}}
  \int_{[0,2\pi]^{d-1}}\frac{1}{h(\theta^1,\theta')-E}d\theta'
\end{equation}
is not constant. We used the notation
$\theta=(\theta_1,\theta_2)=(\theta^1,\theta')$.\\
Recall from Remark~\ref{rem:1} that $h(\theta)=h_0(G'\cdot\theta)$
where $G'\in\text{GSL}(\Z)$ and $h_0$ is defined in~\eqref{eq:7}. So,
the $n$-th Fourier coefficient of $J$ is given by
\begin{equation*}
  \begin{split}
    \hat J_n&=\frac{1}{(2\pi)^d}\int_{[0,2\pi]}
    \int_{[0,2\pi]^{d-1}}\frac{e^{in\theta^1}}{h(\theta^1,\theta')-E}
    d\theta'd\theta^1=\frac{1}{(2\pi)^d}
    \int_{[0,2\pi]^d}\frac{e^{in\theta^1}}{h_0(G'\cdot\theta)-E}d\theta\\
    &=\frac{1}{(2\pi)^d}
    \int_{[0,2\pi]^d}\frac{e^{in(G'^{-1}\cdot\,\theta)^1}}{h_0(\theta)-E}
    d\theta=\frac{e^{in(G'^{-1}\cdot\,\theta_\pi)^1}}{(2\pi)^d}
    \int_{[-\pi,\pi]^d}\frac{e^{in(G'^{-1}\cdot\,\theta)^1}}{-h_0(\theta)-E}
    d\theta
  \end{split}
\end{equation*}
where $(G'^{-1}\cdot\,\theta)^1$ denotes the first coordinate of the
vector $G'^{-1}\cdot\,\theta$, and $\theta_\pi$, the vector
$(\pi,\dots,\pi)$ in $\R^d$. So to prove that $\hat J_n$ does not
vanish for any $n$ which implies that $J$ is not constant, it suffices
to prove that the Fourier coefficients of $(h_0(\theta)-E)^{-1}$ do
not vanish. This is a consequence of the Neuman expansion
\begin{equation*}
  \frac{1}{-h_0(\theta)-E}=\frac{-1}{E}\sum_{k\geq0}
  \left(\frac{h_0(\theta)}{-E}\right)^k.
\end{equation*}
Indeed, the $n$-th Fourier coefficient in each of the terms of order
$k$ larger than $n$ in this series is positive : it is easily seen as
$-E>0$ and the multiplication operator $(h_0)^n$ is unitarily
equivalent through Fourier transformation to $(-\frac12\Delta)^n$; so
the Fourier coefficients of $(h_0)^n$ are the entries of the zeroth
row of the matrix $(-\frac12\Delta)^n$ and, the $n$ first super- and
sub-diagonals of this convolution matrix are positive.

%
%% \bibliographystyle{plain}
%% \bibliography{biblio}
%
  \def\cprime{$'$} \def\cydot{\leavevmode\raise.4ex\hbox{.}}

\end{document}